%% file: main.tex
\documentclass[5p,times,twocolumn]{elsarticle}
\usepackage{physics}
\usepackage{tabularx}
\pdfoutput=1
\DeclareUnicodeCharacter{2212}{-} 

\usepackage{siunitx}
\usepackage{xcolor}
\usepackage{hyperref}
\hypersetup{hidelinks}
\usepackage{csquotes}
\usepackage{amssymb}
\usepackage{pifont}
\newcommand{\cmark}{\ding{51}}%
\newcommand{\xmark}{\ding{55}}%
\usepackage{acro}
\include{sections/00_acro}

\graphicspath{{figures/}}
\usepackage{lipsum}
\usepackage{booktabs}
\usepackage{float}
\usepackage{comment}
\usepackage{subcaption}
\usepackage{framed} 
\usepackage{multicol} 
\usepackage{nomencl} 
\usepackage{todonotes}
\usepackage{tikz}
\usepackage{pgfplots}
\pgfplotsset{width=7.5cm}
\usetikzlibrary{plotmarks}

\usepackage{makecell}

\allowdisplaybreaks
\newcommand\vn[1]{\mathrm{#1}} 

\usepackage{xcolor}
\usepackage{hyperref}

\definecolor{myColor}{RGB}{0, 0.0, 0} 

\hypersetup{
    colorlinks=true,
    linkcolor=myColor,
    citecolor=myColor,
    urlcolor=myColor,
}

\journal{Applied Energy}
\makeatletter
\def\ps@pprintTitle{
  \let\@oddhead\@empty
  \let\@evenhead\@empty
  \def\@oddfoot{}%
  \let\@evenfoot\@oddfoot
}
\makeatother

\begin{document}

\begin{frontmatter}

\title{Pareto local search for a multi-objective demand response problem in residential areas with heat pumps and electric vehicles}

\author[1]{Thomas Dengiz}
\author[2]{Andrea Raith}
\author[1]{Max Kleinebrahm}
\author[1]{Jonathan Vogl}
\author[1]{Wolf Fichtner}

\address[1]{Karlsruhe Institute of Technology, Institute for Industrial Production, Karlsruhe, Germany}
\address[2]{University of Auckland, Department of Engineering Science, Auckland, New Zealand}

\begin{abstract}
\input{sections/00_abstract}
\end{abstract}

\begin{keyword}
Multi-objective optimization \sep demand response \sep Pareto local search  \sep reinforcement learning  \sep residential area
\end{keyword}

\end{frontmatter}


\input{sections/00_nomecl}
\input{sections/01_intro}

\input{sections/02_related_work}

\input{sections/03_optimization_problem_for_the_residential_area}
\input{sections/04_pareto_local_search}
\input{sections/05_results}
\input{sections/06_conclusion}

\input{sections/ending}

\thispagestyle{empty} 


 \bibliographystyle{elsarticle-num} 
 \bibliography{refs}





\end{document}

%% file: sections/00_acro.tex
\DeclareAcronym{EV}{
    short = EV,
    long  = Electric Vehicle,
}

\DeclareAcronym{DHW}{
    short = DHW,
    long  = Domestic hot water,
}

\DeclareAcronym{PLS}{
    short = PLS,
    long  = Pareto local search,
}

\DeclareAcronym{PALSS}{
    short = PALSS,
    long  = Pareto local search for load shifting,
}

\DeclareAcronym{RELAPALSS}{
    short = RELAPALSS,
    long  = Reinforcement learning assisted pareto local search,
}

\DeclareAcronym{MILP}{
    short = MILP,
    long  = Mixed Integer Linear Programming,
}

\DeclareAcronym{RL}{
    short = RL,
    long  = Reinforcement Learning,
}

\DeclareAcronym{NSGA-II}{
    short = NSGA-II,
    long  = Non-dominated Sorting Genetic Algorithm,
}

\DeclareAcronym{SPEA-II}{
    short = SPEA-II,
    long  = Strength Pareto Evolutionary Algorithm,
}

\DeclareAcronym{GD}{
    short = GD,
    long  = Generational Distance,
}

\DeclareAcronym{HV}{
    short = HV,
    long  = Hypervolume,
}

\DeclareAcronym{BT}{
    short = BT,
    long  = Building type,
}

%% file: sections/00_abstract.tex
\acresetall
In future energy systems characterized by significant shares of fluctuating renewable energy sources, there is a need for a fundamental change in electricity consumption. The energy system requires the ability to adapt to the intermittent electricity generation of renewable energy sources. This can be achieved by integrating flexible electrical loads, such as electric heating devices and electric vehicles, in combination with efficient control methods. In this paper, we introduce the Pareto local search method PALSS with heuristic search operations to solve the multi-objective optimization problem of a residential area with different types of flexible loads. PALSS shifts the flexible electricity load with the objective of minimizing the electricity cost and peak load while maintaining the inhabitants' comfort in favorable ranges. Further, we include reinforcement learning into the heuristic search operations in the approach RELAPALSS and use the dichotomous method for obtaining all Pareto-optimal solutions of the multi-objective optimization problem with conflicting goals. The methods are evaluated in simulations with different configurations of the residential area. The results show that PALSS and RELAPALSS strongly outperform the two multi-objective evolutionary algorithms NSGA-II and SPEA-II from the literature and the conventional control approach. The inclusion of reinforcement learning in RELAPALSS leads to additional improvements. Our study reveals the need for multi-objective optimization methods to utilize renewable energy sources in residential areas. 

%% file: sections/00_nomecl.tex
\setlength{\nomitemsep}{-\parskip} 
\makenomenclature
\renewcommand*\nompreamble{\begin{multicols}{2}}
\renewcommand*\nompostamble{\end{multicols}}

\ExplSyntaxOn
\NewExpandableDocumentCommand{\strcase}{mm}
 {
  \str_case:nn { #1 } { #2 }
 }
\ExplSyntaxOff

\renewcommand\nomgroup[1]{%
  \item[\bfseries
    \strcase{#1}{
      {P}{Parameters}
      {A}{Acronyms}
      {V}{Variables}
      {I} {Indices}
    }%
  ]%
}

\begin{table*}[p]
  \begin{framed}
    \nomenclature[A]{\acs{DHW}}{\acl{DHW}}
    \nomenclature[A]{\acs{EV}}{\acl{EV}}
    \nomenclature[A]{\acs{PLS}}{\acl{PLS}}
    \nomenclature[A]{\acs{PALSS}}{\acl{PALSS}}
    \nomenclature[A]{\acs{RELAPALSS}}{\acl{RELAPALSS}}
    \nomenclature[A]{\acs{MILP}}{\acl{MILP}}
    \nomenclature[A]{\acs{RL}}{\acl{RL}}
    \nomenclature[A]{\acs{HV}}{\acl{HV}}
    \nomenclature[A]{\acs{GD}}{\acl{GD}}
    \nomenclature[A]{\acs{NSGA-II}}{\acl{NSGA-II}}
    \nomenclature[A]{\acs{SPEA-II}}{\acl{SPEA-II}}
    \nomenclature[A]{\acs{BT}}{\acl{BT}}

%
    \nomenclature[V]{$T^{\vn{Building}}_{t,b}$}{temperature of the building}
    \nomenclature[V]{$P^{Peak}$}{electrical peak load}
    \nomenclature[V]{$V^{\vn{DHW}}_{t,b}$}{usable volume for DHW of the hot water tank}
    \nomenclature[V]{$Q_{t,b}^{\vn{SH}}$}{heating energy of the heat pump for space heating}
    \nomenclature[V]{$Q_{t,b}^{\vn{DHW}}$}{heating energy of the heat pump for DHW}
    \nomenclature[V]{$x_{t,b}$}{modulation degree of the heat pump for space heating}
    \nomenclature[V]{$y_{t,b}$}{modulation degree of the heat pump for DHW}
    \nomenclature[V]{$h_{t,b}^{\vn{runSH}}$}{binary variable indicating if the heat pump is running for space heating}
    \nomenclature[V]{$h_{t,b}^{\vn{runDHW}}$}{binary variable indicating if the heat pump is running for DHW}

    \nomenclature[V]{$h_{t,b}^{\vn{switchedOff}}$}{binary variable indicating if the heat pump is switched off}
    \nomenclature[V]{$P_{t,b}^{\vn{EV}}$}{charging power of the EV}
    \nomenclature[V]{$\vn{SOC}_{t,b}$}{state of charge of the EV}
    \nomenclature[V]{$P_{i}^{\text{PriceShift}\%}$}{share of shifted load using the Price-Shift-Operator}
    \nomenclature[V]{$P_{i}^{\text{PeakShift}\%}$}{share of shifted load using the Peak-Shift-Operator}

%
    \nomenclature[P]{$T^{\max}_{b}$}{maximum building temperature}
    \nomenclature[P]{$T^{\min}_{b}$}{minimum building temperature}
    \nomenclature[P]{$p_t$}{time-variable electricity price}
    \nomenclature[P]{$V^{\vn{DHWmin}}_{b}$}{minimum volume of the hot water tank}
    \nomenclature[P]{$V^{\vn{DHWmax}}_{b}$}{maximum volume of the hot water tank}
    \nomenclature[P]{$d^{\vn{Temperature}}$}{factor for controlling the allowed deviation from the initial temperature}
    \nomenclature[P]{$d^{\vn{DHW}}$}{factor for controlling the allowed deviation from the initial DHW volume}
    \nomenclature[P]{$Q_{t,b}^{\vn{DemandSH}}$}{demand for space heating}
    \nomenclature[P]{$Q_{t,b}^{\vn{DemandDHW}}$}{demand for DHW}
    \nomenclature[P]{$Q_{t,b}^{\vn{LossesSH}}$}{energy losses of the heating system}
    \nomenclature[P]{$Q_{t,b}^{\vn{LossesDHW}}$}{energy losses of the hot water tank}

    \nomenclature[P]{$P^{\vn{HP}}_b$}{maximum electrical power of the heat pump}
    \nomenclature[P]{$\vn{COP}_{t,b}$}{coefficient of performance of the heat pump}
    \nomenclature[P]{$k_b$}{maximum number of heat pump starts}
    \nomenclature[P]{${\vn{mod}}^{\min}$}{minimum modulation degree of the heat pump}
    \nomenclature[P]{$P_{t,b}^{\vn{EVMax}}$}{maximum power of the EV charging station}
    \nomenclature[P]{$a_{t,b}$}{availability parameter of the EV}
    \nomenclature[P]{$\eta_{b}$}{charging efficiency of the EV}
    \nomenclature[P]{$P_{t,b}^{\vn{EVDrive}}$}{power consumption of the EV while driving}
    \nomenclature[P]{$C^{\vn{EV}}_b$}{energy capacity of the EV's battery}
    \nomenclature[P]{$P_{t,b}^{\vn{el}}$}{electrical power of the inflexible devices}

    \nomenclature[P]{$V^{\vn{UFH}}_b$}{volume of the underfloor heating system}
    \nomenclature[P]{$T^{\vn{DHW}}$}{temperature of the domestic hot water}
    \nomenclature[P]{$\rho^{\vn{Concrete}}$}{desitiy of concrete}
    \nomenclature[P]{$c^{\vn{Concrete}}$}{specific heat capacity of concrete}

    \nomenclature[I]{$b$}{building}
    \nomenclature[I]{$i$}{iteration}
    \nomenclature[I]{$t$}{time slot}
    \nomenclature[I]{$Z$}{total number of time slots}
    \nomenclature[I]{$B$}{total number of buildings}
    \printnomenclature
  \end{framed}
\end{table*}

%% file: sections/01_intro.tex
\section{Introduction}
\label{sec:introduction}
Future energy systems require flexible electricity loads to cope with the intermittent electricity generation by renewable energy sources. In residential areas, electric heating systems (like heat pumps) coupled to thermal storage, and electric vehicles (EV), are especially suitable for demand response by adjusting their electricity consumption based on an external signal \cite{Shao2011, Patteeuw2015}. A price signal tied to electricity market prices is typically used to control electricity consumption \cite{SHARIATZADEH2015}. The problem with using a price or availability signal for renewable energy sources is that all buildings tend to react to the lowest price at the same time \cite{KUHNBACH2021}. This leads to new peak loads in the electricity grid that might harm the transformers or the transmission lines. 

Thus, the resulting optimization problem for controlling flexible devices in a residential area has multiple conflicting goals. Next to the minimization of the electricity cost, the reduction of the peak loads in the local grid is an important objective. Further, considering the inhabitants' thermal comfort is crucial for shifting the operation of the electrical heating devices. As solving the (multi-objective) scheduling problem is NP-hard \cite{ULLMAN1975}, heuristic local search methods based on Pareto-optimality have been successfully used in several applications of multi-objective optimization \cite{Luo2022, Paquete2004}. 

Hence, we define a  Pareto local search method for solving the resulting multi-objective demand response optimization problem of a residential area. We include different building types (BT) with varying flexibility options to adjust the electricity consumption based on an external price signal. The goal is to minimize the electricity costs while simultaneously reducing the peak load and maintaining the thermal comfort of the inhabitants in acceptable ranges. 

The remainder of the paper is organized as follows: In Sec.~\ref{sec_related_word_and_contribution}, we describe the relevant literature and explain the main contribution of this paper. We define the multi-objective optimization problem in Sec.~\ref{sec:Optimization problem for the residential area} and introduce our novel Pareto local search approach and an exact method from the literature for solving the problem in Sec.~\ref{sec:Methods for solving the multi-objective optimization problem}. The results of our experiments are presented in Sec.~\ref{sec:Results}. The paper ends with a conclusion and outlook in Sec.~\ref{sec:Conclusion}.

%% file: sections/02_related_work.tex
\section{Related work and contribution}
\label{sec_related_word_and_contribution}

\subsection{Related work}
\label{subsec_Related work}

\begin{table*} 
    \centering
    \caption{Comparison of relevant papers from the literature}
    \begin{tabular}{|l|c|c|c|c|c|}
        \hline
        & \makecell{RL with domain \\ knowledge}  & \makecell{Multi-objective \\ optimization} & \makecell{Pareto \\ local search} & \makecell{Compared to \\ metaheuristics } & \makecell{Compared to \\ exact methods } \\
        \hline
        Pinto et al., 2021 \cite{PINTO20211}  & \xmark  & (\cmark) & \xmark & \xmark & \xmark\\
        \hline
         Peirelinck et al., 2024 \cite{PEIRELINCK2024}  & \cmark & (\cmark) & \xmark & \xmark & \xmark\\
         \hline
         Pinto et al., 2021 \cite{PINTO20211_2}  & \cmark &  (\cmark) &  \xmark & \xmark & \xmark\\
         \hline
         Shakouri et al., 2017 \cite{SHAKOURIG2017}  & \xmark  & (\cmark) & \xmark & \xmark & \cmark\\
         \hline
         Tang et al., 2022 \cite{Tang2022} & \xmark & (\cmark) & \xmark & \xmark & \xmark \\
         \hline
         Dutra et al., 2019 \cite{Dutra2019} & \xmark  & (\cmark) & \xmark & \xmark & \cmark \\
         \hline
         Khalid et al., 2018 \cite{Khalid2018} & \xmark  & (\cmark) & (\cmark) & \cmark & \xmark \\
         \hline
         Wynn et al., 2022 \cite{Wynn2022} & \xmark  & \cmark & \xmark & \cmark & \xmark \\
         \hline
         Terlouw et al., 2019 \cite{Terlouw2019} & \xmark  & \cmark & \xmark & \xmark & (\cmark) \\
         \hline
          Song et al., 2022 \cite{Song2022} & \xmark  & (\cmark) & \xmark & \xmark & (\cmark) \\
         \hline
        Wang et al., 2020 \cite{Wang2020} & \xmark & \cmark & \xmark & \cmark & \xmark \\
         \hline
        Kazmi et al., 2019 \cite{Kazmi2019} & \cmark & \xmark & \xmark & \xmark & \xmark \\
         \hline
        Wei et al., 2015 \cite{Wei2015} & \xmark &  \cmark & \xmark & \cmark & \xmark \\
         \hline
        Chiu et al., 2020 \cite{Chiu2020} & \xmark  & \cmark & \cmark & \cmark & \xmark \\
         \hline
        Present work  & \cmark &  \cmark & \cmark & \cmark & \cmark \\
         \hline
    \end{tabular}
    \label{tab_literature}
\end{table*}

Table~\ref{tab_literature} lists the relevant studies from the literature. Pinto et al.~\cite{PINTO20211} use (Reinforcement Learning) RL to optimize four buildings with heat pumps and domestic hot water (DHW). They aim to minimize the energy costs and the peak load. Peirelinck et al.~\cite{PEIRELINCK2024} include domain knowledge into their RL application to control the DHW heating of buildings to minimize the electricity cost by having a tariff that depends on the peak load. Pinto et al.~\cite{PINTO20211_2} combine a Long-Short-Term-Memory (LSTM) with RL to control heat pumps. They use the LSTM to learn the building dynamics for evaluation from a building simulation tool. Shakouri et al.~\cite{SHAKOURIG2017} define a multi-objective optimization problem for minimizing the costs and peak load in an intelligent grid by shifting the electricity consumption of household devices. They solve the problem with Goal programming. In \cite{Tang2022}, Tang et al.~use RL to operate an integrated energy system with electricity, gas, and heating networks. Their objectives also include the minimization of costs and peak load. Dutra et al.~\cite{Dutra2019} introduce a decentralized heuristic for demand response of an aggregator. The aggregator controls multiple flexible loads such that the costs and the peak load are reduced. Khalid et al.~2018 \cite{Khalid2018} define a novel local search method based on two metaheuristics (genetic algorithm and hybrid bacterial foraging). Their flexible devices include air conditioners (AC), water heaters, washing machines, and dishwashers. 

Wynn et al.~\cite{Wynn2022} use a multi-objective version of the metaheuristic grey wolf optimization applied for scheduling flexible resources in a microgrid. Next to minimizing the costs and the peak load, they consider consumers' dissatisfaction in their optimization problem. Terlouw et al.~\cite{Terlouw2019} define a multi-objective optimization problem for an aggregator with different battery technologies for a whole community. The objectives comprise minimizing costs, CO$_2$-emissions, and the peak load. In the study \cite{Song2022}, Song et al.~solve a day-ahead battery scheduling problem for residential buildings and include the goal of costs, peak load, and consumer satisfaction in the objective function. Wang et al.~\cite{Wang2020} use the multi-objective genetic algorithm NSGA-II for the optimization of a combined cooling, heating, and power system. The goals are the reduction of CO$_2$-emissions, costs, energy usage, and the enhancement of grid integration. Kazmi et al.~\cite{Kazmi2019} train an RL agent for thermostatically controlled loads (heat pump, DHW). They include domain knowledge into RL by adding additional training features derived from thermodynamic laws. The goal is to minimize energy consumption in residential areas. Wei et al.~\cite{Wei2015} use multi-objective particle swarm optimization to optimize the operation of a heating-ventilation and air conditioning (HVAC) system. The goals include minimizing the energy consumption, CO$_2$-emissions and the temperature of buildings while maximizing the indoor air quality. Chiu et al.~\cite{Chiu2020} introduce a Pareto local search approach based on evolutionary algorithms for demand response in a residential area. The Pareto local search controls the heat pump, AC, EV, and shiftable devices to minimize the energy costs and the peak load. 

While every study considers multiple objectives, most use the weighted sum approach to transform the multi-dimensional objective space into a one-dimensional space (ticks in brackets). Only \cite{Wynn2022, Terlouw2019, Wang2020, Wei2015, Chiu2020} optimize in a multi-objective space and generate trade-off (Pareto-optimal) solutions. However, none of those studies compares their introduced approach to an exact method for multi-objective optimization that systematically generates the true Pareto-front containing all Pareto-optimal solutions. Only Chiu et al.~\cite{Chiu2020} introduce a Pareto local search approach that heuristically tries to approximate the true Pareto-front using local search methods. While several studies use RL \cite{PINTO20211, PEIRELINCK2024, PINTO20211_2, Tang2022, Kazmi2019}, only \cite{PEIRELINCK2024, PINTO20211_2,  Kazmi2019} include domain knowledge into the decision making of their trained agent. 

\subsection{Contribution}
\label{subsec_Contribution}
To the best of our knowledge, our study is the only one that introduces a Pareto local search method for shifting flexible electricity loads in residential areas in a multi-dimensional objective space. When dealing with multi-objective optimization problems, approaches for finding Pareto-optimal aim to find optimal trade-off solutions for decision-making. This feature of heuristic control approaches for demand response in residential areas has not been adequately explored in the literature. We also extend our approach by integrating RL into the decision-making and evaluating the novel approaches. In summary, our paper has the following three unique contributions:

\begin{itemize}
  \item We introduce a novel approach for Pareto local search for demand response in residential areas that utilizes heuristic search operations.
  \item We include a model-free RL approach into the Pareto local search's decision-making and thus include domain knowledge into the training of the RL agent. 
  \item For the evaluation, we solve the multi-objective optimization problem for a residential area using two state-of-the-art metaheuristics and an exact method that can guarantee to find the whole Pareto-front. Further, we compare them to an existing conventional control approach that is used most often nowadays in building heating systems. 
\end{itemize}

%% file: sections/03_optimization_problem_for_the_residential_area.tex
\section{Multi-objective demand response optimization problem for the residential area}
\label{sec:Optimization problem for the residential area}

\begin{figure}[htb]
    \centering
    \begin{subfigure}[h]{0.49\textwidth}
        \centering
        \includegraphics[width=\textwidth]{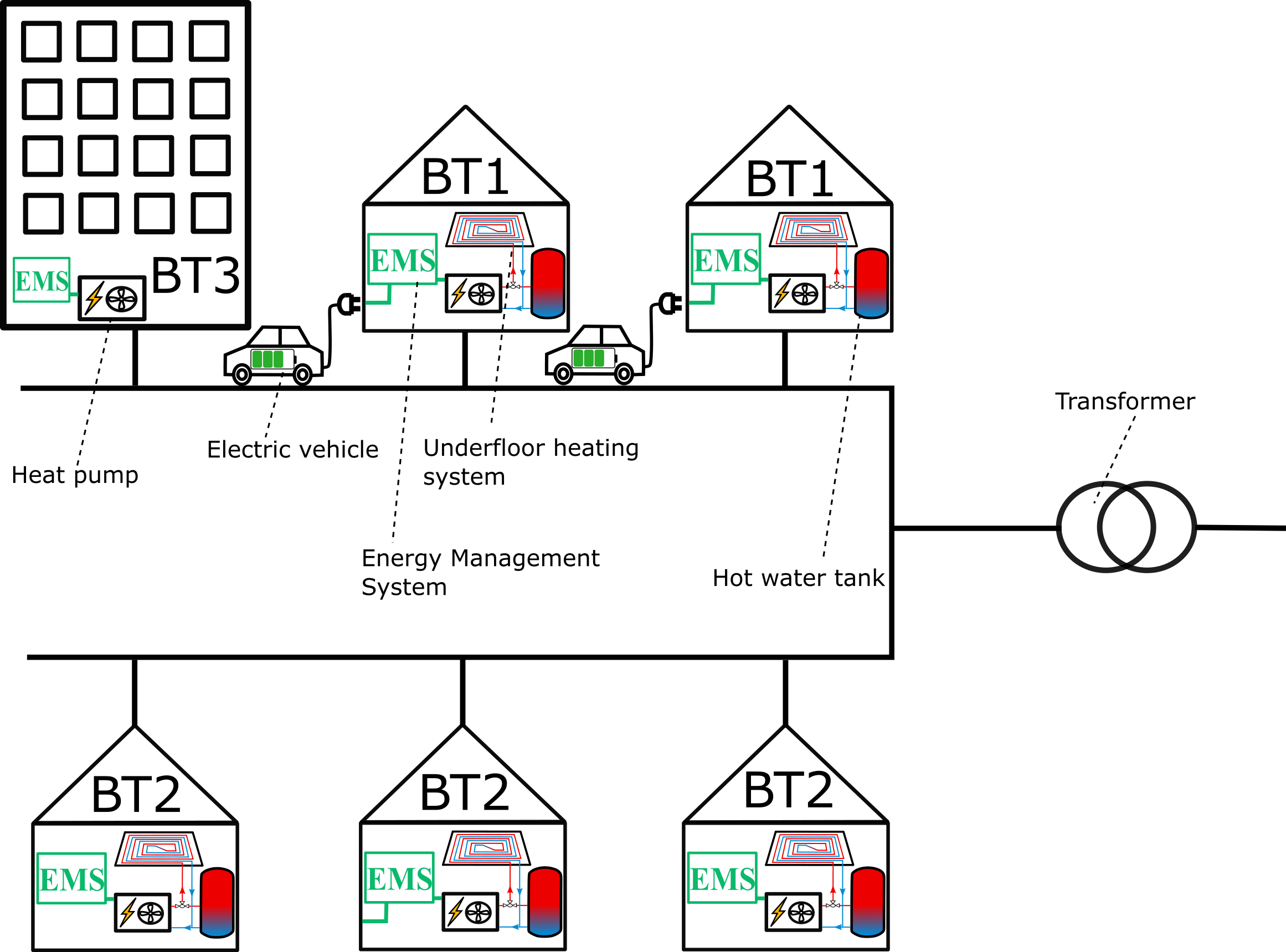}

    \end{subfigure}
   \caption{Residential area with three different building types}
    \label{fig:Residential_Area_Paper_PLS}
\end{figure}

In this section, we formulate the multi-objective demand response optimization problem for the residential area, which consists of three building types visualized in Fig.~\ref{fig:Residential_Area_Paper_PLS}. BT1 is a single-family building that uses a heat pump for space heating and domestic hot water. An underfloor heating system and a hot water tank serve as thermal storage. Further, the inhabitants charge the EV at home. BT2 is similar to BT1, but does not use an EV. BT3 is a multi-family building that uses a heat pump for space heating but not for DHW supply (DHW requires higher temperatures which reduces the efficiency of heat pumps). 

The primary goals for a centralized controller are to minimize the total electricity costs with a time-variable electricity tariff $p_t$ and to minimize the peak load $P^{Peak}$ in the residential area.

\begin{equation}\label{eq:objective_function}
    \min (\vn{\vn{Cost}}, P^{\vn{Peak}}) 
\end{equation}
subject to:
\begin{align}
    &T^{\min}_{b} \leq T^{\vn{Building}}_{t,b} \leq T^{\max}_{b} \quad \forall t, b  
       \label{eq:temperature_balance} \\
    &V^{\vn{DHWmin}}_{b} \leq V^{\vn{DHW}}_{t,b} \leq V^{\vn{DHWmax}}_{b} \quad \forall t, b  
       \label{eq:DHW_balance} \\
    & T^{\vn{Building}}_{Z,b} \geq T^{\vn{Building}}_{1,b} \cdot d^{\vn{Temperature}} \quad \forall b  
       \label{eq:Tempeature_End_Value} \\
    & V^{\vn{DHW}}_{Z,b} \geq V^{\vn{DHW}}_{1,b} \cdot d^{\vn{DHW}} \quad \forall b  
       \label{eq:DHW_End_Value} \\
    &T^{\vn{Building}}_{t,b} = \notag \\
    &\quad T^{\vn{Building}}_{t-1,b} 
           + \frac{Q_{t,b}^{\vn{SH}} - Q_{t,b}^{\vn{DemandSH}} -  Q_{t,b}^{\vn{LossesSH}}}%
                  {V^{\vn{UFH}}_b \cdot  \rho^{\vn{Concrete}} \cdot c^{\vn{Concrete}}} \quad \forall t \neq 1, \forall b  
        \label{eq:temperature_difference_equation}\\
    &V^{\text{DHW}}_{t,b} = \notag \\
    &\quad V^{\text{DHW}}_{t-1, b} 
           + \frac{Q_{t,b}^{\text{DHW}} - Q_{t,b}^{\text{DemandDHW}} - Q_{t,b}^{\text{LossesDHW}}}%
                  {T^{\text{DHW}} \cdot  \rho^{\text{Water}} \cdot c^{\text{Water}}} \quad \forall t \neq 1, \forall b  
           \label{eq:volume_difference_equation} \\
    &Q_{t,b}^{\vn{SH}} = x_{t,b} \cdot P^{\vn{HP}}_b  \cdot \vn{COP}_{t,b} \cdot \Delta t \quad \forall t, b   
       \label{eq:heating_energy_HP_SH} \\
    &Q_{t,b}^{\vn{DHW}} = y_{t,b} \cdot P^{\vn{HP}}_b  \cdot \vn{COP}_{t,b} \cdot \Delta t  \quad \forall t, b  
       \label{eq:heating_energy_HP_DHW} \\
    &h_{t-1,b}^{\vn{runSH}} + h_{t-1,b}^{\vn{runDHW}} 
       \leq h_{t,b}^{\vn{runSH}} + h_{t,b}^{\vn{runDHW}} + h_{t,b}^{\vn{switchedOff}}  \quad \forall t \neq 1, \forall b  
       \label{eq:numer_of_switch_offs_1} \\
    &h_{t,b}^{\vn{runSH}} + h_{t,b}^{\vn{runDHW}} \leq 1  \quad \forall t, b 
       \label{eq:numer_of_switch_offs_2} \\
    & \sum\nolimits_{t=1}^{Z} h_{t,b}^{\vn{switchedOff}} \leq k_b  \quad \forall b  
       \label{eq:numer_of_switch_offs_3} \\
    &x_{t,b} \leq h_{t,b}^{\vn{runSH}} \quad \forall t, b 
       \label{eq:min_mod_SH_1}   \\
    &x_{t,b} \geq h_{t,b}^{\vn{runSH}} \cdot {\vn{mod}}^{\min} \quad \forall t, b
       \label{eq:min_mod_SH_2}   \\
    &y_{t,b} \leq h_{t,b}^{\vn{runDHW}} \quad \forall t, b
       \label{eq:min_mod_DHW_1} \\
    &y_{t,b} \geq h_{t,b}^{\vn{runDHW}} \cdot {\vn{mod}}^{\min} \quad \forall t, b
       \label{eq:min_mod_DHW_2} \\
    &P_{t,b}^{\vn{EV}} \leq P_{t,b}^{\vn{EVMax}} \cdot a_{t,b} \quad \forall t, b
       \label{eq:max_power_EV} \\
    &0 \leq \vn{SOC}_{t,b} \leq 1 \quad \forall t, b 
       \label{eq:SOC_Balance} \\
    & \vn{SOC}_{Z,b} \geq \vn{SOC}_{1,b} \cdot d^{\vn{SOC}} \quad \forall b
          \label{eq:SOC_End_Value} \\
    & \vn{SOC}_{t,b} = \notag \\
    & \quad \vn{SOC}_{t-1,b} 
           + \frac{P_{t,b}^{\vn{EV}} \cdot \eta_{b} \cdot \Delta t - P_{t,b}^{\vn{EVDrive}} \cdot \Delta t}%
                 {C^{\vn{EV}}_b}, \quad \forall t \neq 1, \forall b 
       \label{eq:soc_difference_equation} \\
    &\vn{Cost} = \sum\nolimits_{t=1}^{Z} \sum\nolimits_{b=1}^{B} 
                 ((x_{t,b} + y_{t,b}) \cdot P^{\vn{HP}}_b + P_{t,b}^{\vn{EV}} + P_{t,b}^{\vn{el}}) 
                 \cdot \Delta t \cdot p_t   
       \label{eq:objective_cost} \\
    &P^{\vn{Peak}} =  \max_{t} \Bigl\{ 
       \sum\nolimits_{b=1}^{B} \bigl((x_{t,b} + y_{t,b}) \cdot P^{\vn{HP}}_b + P_{t,b}^{\vn{EV}} + P_{t,b}^{\vn{el}}\bigr)\Bigr\} \label{eq:peak} \\
    &\begin{aligned}
       &x_{t,b} \in [0,1], 
          && y_{t,b} \in [0,1], \\
       &h_{t,b}^{\vn{runSH}} \in \{0,1\}, 
          && h_{t,b}^{\vn{runDHW}} \in \{0,1\}, 
          && h_{t,b}^{\vn{switchedOff}} \in \{0,1\}
     \end{aligned} 
\end{align}

Eq.~\ref{eq:temperature_balance} makes sure that the temperature $T^{\vn{Building}}_{t,b}$ for every building $b$ is always between a lower temperature limit $T^{\min}_{b}$ and an upper temperature limit $T^{\max}_{b}$ during every time slot $t$. In analogy, Eq.~\ref{eq:DHW_balance} forces the volume of the hot water tank $V^{\vn{DHW}}_{t,b}$ to be between the corresponding minimum $^{\vn{DHWmin}}_{b}$ and maximum volume $V^{\vn{DHWmax}}_{b}$. To ensure that the values of the building temperature and the usable volume of the hot water tank are not too far away from their initial values, we add Eq.~\ref{eq:Tempeature_End_Value} and Eq.~\ref{eq:DHW_End_Value}. These equations include the two parameters $d^{\vn{Temperature}}$ and $d^{\vn{DHW}}$ whose values are between 0 and 1. Thus, we can control the deviation from the allowed initial values. 

We define the temperature difference equation in Eq.~\ref{eq:temperature_difference_equation}. The temperature of each building $b$ at time $t$ is equal to its temperature from the previous time slot plus the energetic difference divided by the volume of the underfloor heating system $V^{\vn{UFH}}$, the density $\rho^{\vn{Concrete}}$ and the specific heat capacity of concrete $c^{\vn{Concrete}}$. The heat pump's heating energy for space heating $Q_{t,b}^{\vn{SH}}$ increases the temperature of the building while the demand for space heating $Q_{t,b}^{\vn{DemandSH}}$ and the losses of the heating system itself $Q_{t,b}^{\vn{LossesSH}}$ decrease it. 
The demand for space heating comprises a time series incorporating the building's transmission and ventilation losses, as well as internal and solar gains. We describe in Sec.~\ref{sec:Scenarios for the analysis} how we generated the time series. In analogy, 
Eq.~\ref{eq:volume_difference_equation} defines the energetic difference equation for the volume of the hot water tank. The heating energy flows for DHW from the heat pump $Q_{t,b}^{\vn{DHW}}$ increase the usable volume of DHW while the demand for DHW $Q_{t,b}^{\vn{DemandDHW}}$ and the hot water tank's standing losses $Q_{t,b}^{\vn{LossesDHW}}$ decrease it. As hot water tanks use water as storage medium, the energetic difference is divided by the temperature $T^{\vn{DHW}}$, density $\rho^{\vn{Water}}$ and specific heat capacity $c^{\vn{Water}}$ of water. 

To calculate the heat pump's heating energy for space heating, Eq.~\ref{eq:heating_energy_HP_SH} multiplies the modulation degree of the heat pump for space heating $x_{t,b}$ with the heat pump's maximum electrical power $P^{\vn{HP}}_b$, the coefficient of performance $\vn{COP}_{t,b}$ and the time resolution $\Delta t$. Likewise, Eq.~\ref{eq:heating_energy_HP_DHW} uses the heat pump's modulation degree for heating the hot water tank $y_{t,b}$ to derive the heating flows to the hot water tank. 

Eq.~\ref{eq:numer_of_switch_offs_1} introduces the binary variables $h_{t,b}^{\vn{runSH}}$ and $h_{t,b}^{\vn{runDHW}}$ that indicate whether the heat pump is running for space heating or DHW at time $t$. Further, the binary variable $h_{t,b}^{\vn{switchedOff}}$ has the value $1$ if the heat pump is being switched off at time slot $t$. As the heat pump cannot heat the building (space heating) and the hot water tank simultaneously, we add Eq.~\ref{eq:numer_of_switch_offs_2}. To ensure that the maximum number of heat pump starts does not exceed $k_b$, Eq.~\ref{eq:numer_of_switch_offs_3} is added. Eq.~\ref{eq:min_mod_SH_1} to Eq.~\ref{eq:min_mod_DHW_2} force the heat pump's modulation degree to be either $0$ or higher than the minimum modulation degree ${\vn{mod}}^{\min}$.

The charging power for the electric vehicle $P_{t,b}^{\vn{EV}}$ should not exceed the maximum power of the charging station $P_{t,b}^{\vn{EVMax}}$. Moreover, the EV can only be charged at a particular time slot $t$ if its binary availability variable $a_{t,b}$ indicates that the EV is at home and thus available for charging. Hence, we introduce Eq.~\ref{eq:max_power_EV} to incorporate these constraints.
Eq.~\ref{eq:SOC_Balance} ensures that the state of charge of the electric vehicle's battery $\vn{SOC}_{t,b}$ is always between $0$ and $1$, while Eq.~\ref{eq:SOC_End_Value} hinders it to be too far away from its initial value at the end of the optimization horizon $Z$. The difference equation for the $\vn{SOC}_{t,b}$ is defined in Eq.~\ref{eq:soc_difference_equation}. The charging power of the EV, adjusted by the charging efficiency $\eta_{b}$, increases the $SOC$ while the required electrical power while driving the EV $P_{t,b}^{\vn{EVDrive}}$ reduces it. The parameter $C^{\vn{EV}}_b$ quantifies the energetic capacity of the battery. 

Eq.~\ref{eq:objective_cost} defines the costs and, thus, the first objective. The total costs are calculated by summing the electrical power consumption of the heat pump, the EV, and the inflexible household appliances $P_{t,b}^{\vn{el}}$. This is multiplied by the time resolution and the current price $p_t$ of a time-variable electricity tariff. For the costs, we sum up over all time slots and buildings. 

To determine the peak load as the second objective, we use Eq.~\ref{eq:peak}, which takes the maximum value of the combined electrical load of all buildings over all time slots. 

While the defined optimization problem is valid for the whole residential area, it must be noted that BT2 and BT3 do not have an EV, and BT3 only uses the heat pump for space heating. For these buildings, the corresponding equations are neglected. The optimization problem is a mixed-integer linear problem (MILP). 

%% file: sections/04_pareto_local_search.tex
\section{Methods for solving the multi-objective optimization problem}
\label{sec:Methods for solving the multi-objective optimization problem}
In this section, we introduce our novel algorithms for shifting electrical loads in a residential area and thereby find Pareto-optimal solutions for the minimization of the peak load and the electricity costs. A solution is Pareto-optimal when there exists no other solution that improves at least one objective without worsening the others.

In Sec.~\ref{subsec:Pareto local search (PASS)}, we explain the Pareto local search approach for demand side management \textit{PALSS}, which constitutes the main contribution of this paper. Next, we describe \textit{RELAPALSS} in Sec.~\ref{subsec_Reinforcement_learning_assisted_PLS}, which uses reinforcement learning to assist \textit{PALSS}. 
In Sec.~\ref{subsec_Dichotomous_method}, we explain the \textit{Dichotomous method}, which is an exact method for multi-objective optimization that we use to obtain a real Pareto-front of the problem that consists of so-called supported Pareto-optimal solutions. We use a conventional control approach for a lower benchmark in our evaluations that we describe in Sec.~\ref{subsec_Conventional_control}.

\subsection{Pareto local search for load shifting (PALSS)}
\label{subsec:Pareto local search (PASS)}

\subsubsection{Pareto local search } 
\label{subsubsec:Pareto_local_search_general}
\begin{figure}[htb]
    \centering
    \begin{subfigure}[h]{0.49\textwidth}
        \centering
        \includegraphics[width=\textwidth]{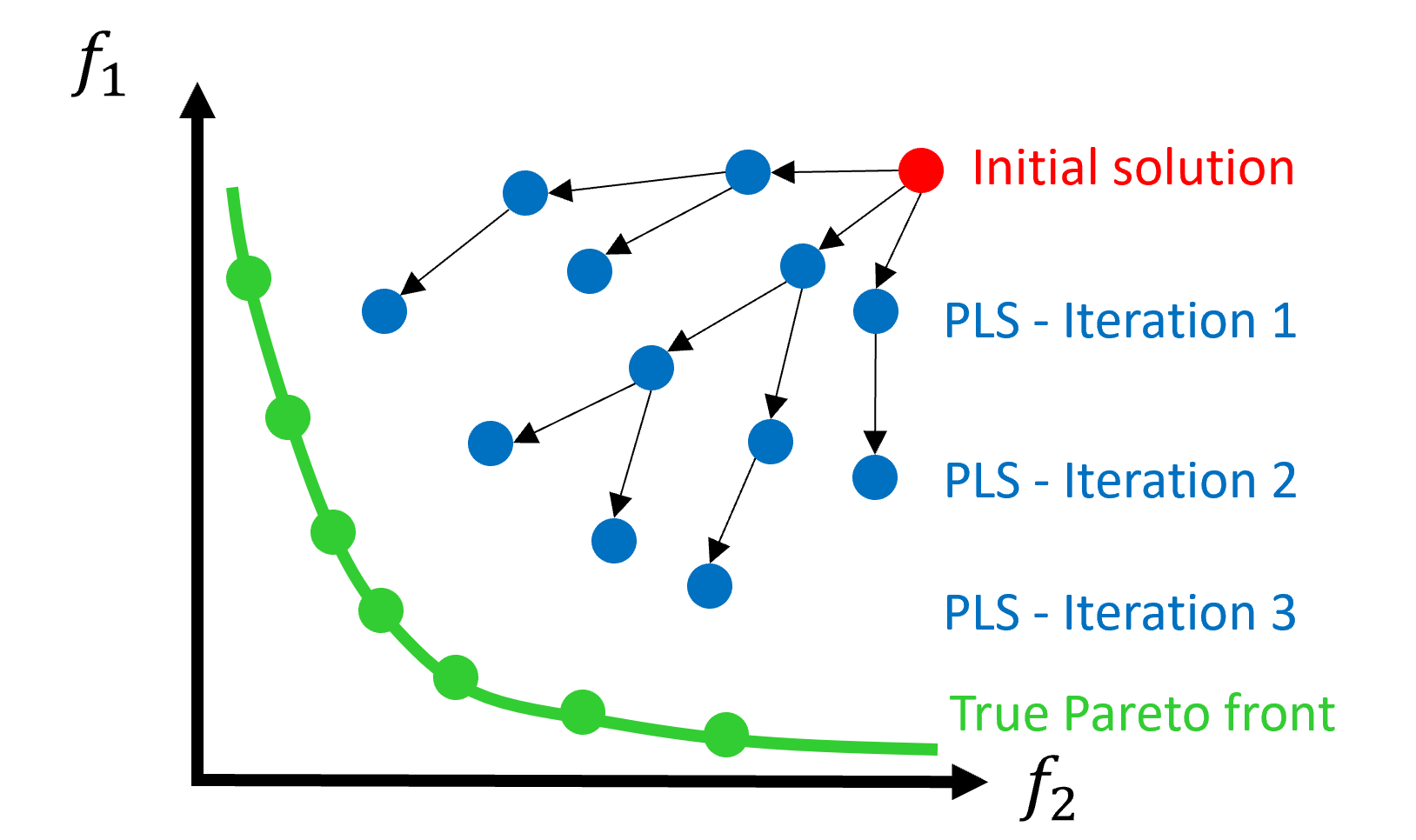}

    \end{subfigure}
   \caption{Solution space of a minimization problem with Pareto local search}
    \label{fig_Solution_Space_Pareto_Local_Search_General}
\end{figure}

Pareto local search (PLS) aims to find a good approximation of the true Pareto-front (that is not known before) in a reasonable time. PLS extends the iterative improvement algorithm, called local search, from single-objective to multi-objective problems for the systematic exploration of the solution space \cite{Paquete2004}. It navigates the neighborhood of solutions within its ve, incorporating new non-dominated solutions while removing dominated ones to refine the ve. Fig.~\ref{fig_Solution_Space_Pareto_Local_Search_General} illustrates the outcomes in the solution space of three PLS iterations for the minimization of two objectives. In many applications, the search starts from a naive solution of the problem that is easy to find.

\begin{figure}[htb]
    \centering
    \begin{subfigure}[h]{0.50\textwidth} %
        \centering
        \includegraphics[width=\textwidth]{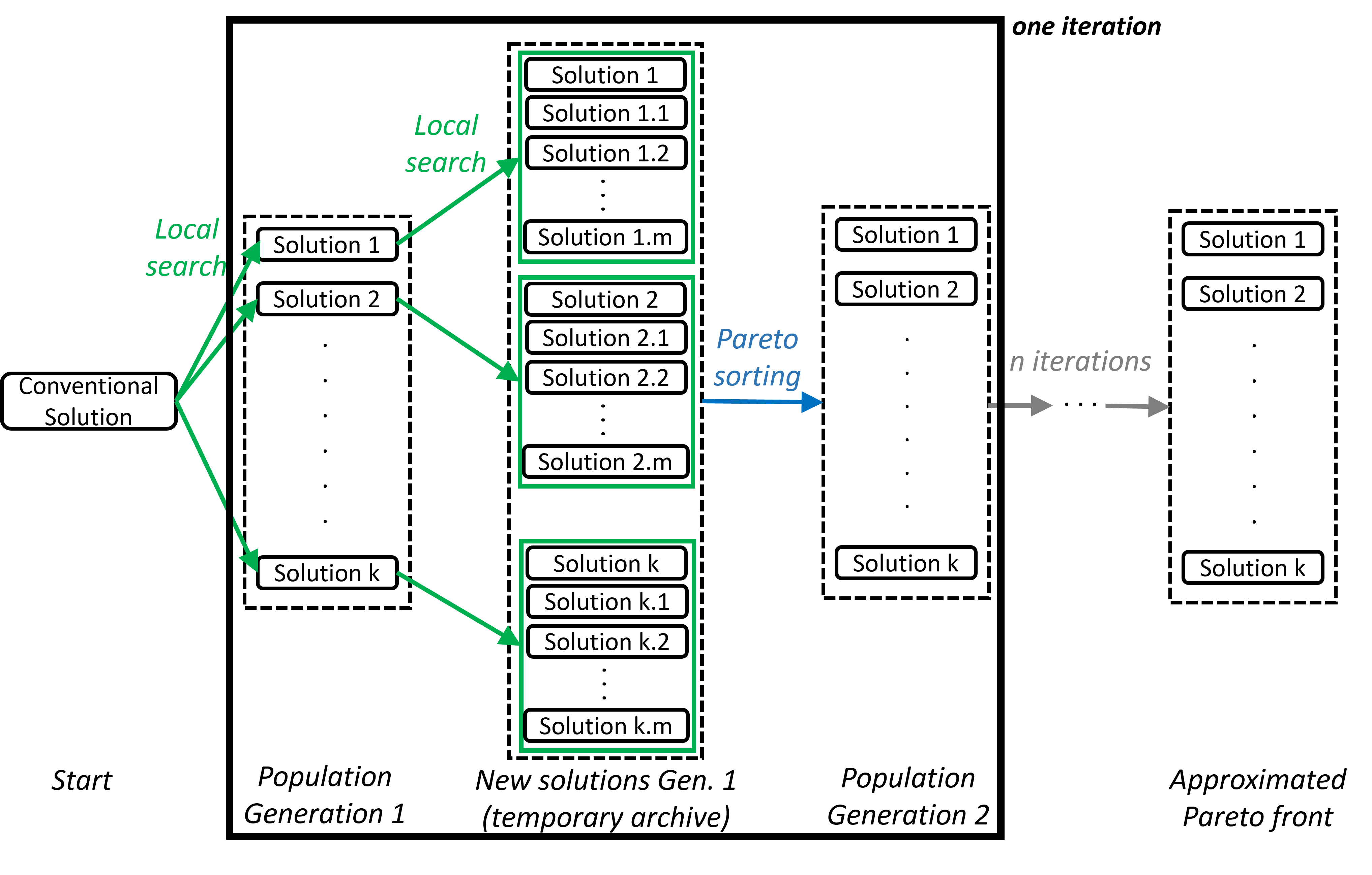}
    \end{subfigure}
    \caption{Schema of PALSS}
    \label{fig_Schema_PALSS}
\end{figure}

Fig.~\ref{fig_Schema_PALSS} depicts the schematic workflow of our developed method \textit{PALSS}. The search starts with a conventional solution (see Sec.~\ref{subsec_Conventional_control}).

\textit{PALSS} generates the population of the first generation by using local search operators (see Sec.~\ref{subsubsec:Local_Search_Operators}) to generate $k$ solutions. Afterward, the same local search operations generate $m$ new solutions from each of the $k$ existing solutions and store each of the $k\cdot m$ solutions in a temporary archive. Next, \textit{PALSS} uses the Pareto sorting approach (see Sec.~\ref{subsubsec_Pareto_sorting}) to filter the solutions in the temporary archive and reduce them again to $k$ solutions. These solutions are Pareto-optimal (based on the solutions in the archive) and form the population of the next iteration. This procedure is repeated for $n$ iterations (in case the algorithm meets no other stopping criterion, such as a time limit or predefined target values of the different goals), and the algorithm outputs the approximated Pareto front.

\subsubsection{Local Search Operators} 
\label{subsubsec:Local_Search_Operators}
\begin{figure}[htb]
    \centering
    \begin{subfigure}[h]{0.49\textwidth}
        \centering
        \includegraphics[width=\textwidth]{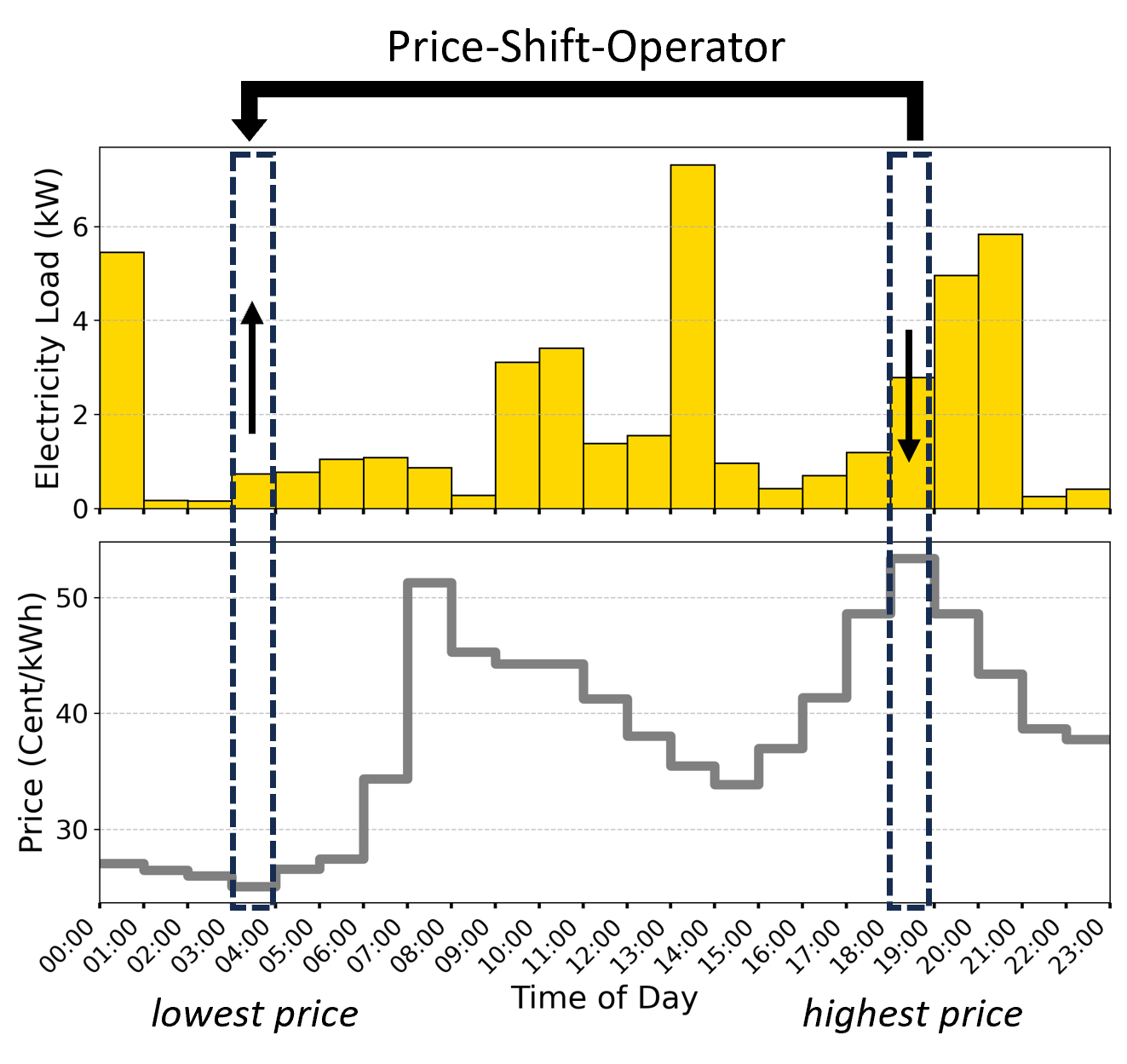}

    \end{subfigure}
   \caption{Price-Shift-Operator}
    \label{fig_Price_Shift_Operator}
\end{figure}

\textit{PALSS} makes use of two shifting operations to adjust the load profiles of individual buildings: the \textit{Price-Shift-Operator} and the \textit{Peak-Shift-Operator}. Both use as input the resulting electrical load profile of the residential area from a solution and the time-variable electricity tariff. The \textit{Price-Shift-Operator} shifts loads from time slots with high electricity prices to time slots with low prices. Fig.~\ref{fig_Price_Shift_Operator} illustrates this operation. Flexible electrical devices (heat pumps, EVs) try to reduce their load during time slots with high prices if possible. In compensation for that, the devices try to increase their electricity consumption during low-price periods.

For the selection of the affected time slots and the amount of shifted loads, we use a heuristic approach. Table~\ref{tab_selection_probabilities} shows the probabilities for selecting the hours with the highest, second highest, third highest prices, etc., during each iteration. The probability of selecting the time slot with the highest and second highest prices is very high initially and slightly decreases in every new iteration. In contrast, the probability of selecting the hour with the third, fourth, and fifth highest price to reduce the load increases in each iteration. The probabilities for selecting the time slot with the lowest value to shift the load to are determined analogously. Using selection probabilities enhances the exploration of the search space. We determined these probabilities after running a large number of experiments. The concrete probabilities depend on the specific use case and scenario. However, the general idea of having high probabilities at the beginning for the highest prices which decrease over time, and analogously, having lower probabilities for the lower prices that slightly increase in every iteration, is useful for all probabilistic demand response heuristics. 

\begin{table}[h]
\centering
\caption{Probabilities of \textit{PALSS} for selecting the highest price hours in each iteration using the \textit{Price-Shift-Operator} (Probabilities for the lowest price hours are equivalent to those for the highest price hours)}
\label{tab_selection_probabilities}
\begin{tabular}{cccccc}
\toprule
\makecell{Iteration} & \makecell{Highest \\ hour 1} & \makecell{Highest \\ hour 2} & \makecell{Highest \\ hour 3} & \makecell{Highest \\ hour 4} & \makecell{Highest \\ hour 5} \\
\midrule
1 & 0.410 & 0.328 & 0.123 & 0.082 & 0.057 \\
2 & 0.393 & 0.311 & 0.139 & 0.098 & 0.059 \\
3 & 0.377 & 0.295 & 0.156 & 0.11 & 0.062 \\
4 & 0.361 & 0.279 & 0.172 & 0.12 & 0.068 \\
5 & 0.344 & 0.262 & 0.189 & 0.135 & 0.07 \\
\bottomrule
\end{tabular}
\end{table}

Next to selecting the time slots for load shifting, the \textit{Price-Shift-Operator} determines the amount of load that is shifted between the time slots. For this purpose, it uses the following uniform distribution in iteration $i$:

\begin{equation}\label{eq_amount_load_shifting}
\Delta P_{i}^{\text{PriceShift}\%} \sim U(20-i, 40-2i), \quad i \in \{1, \ldots, 5\}
\end{equation}

According to Eq.~\ref{eq_amount_load_shifting}, the percentage of shifted loads $P_{i}^{\text{load}\%}$ in iteration $i$ is randomly determined by using a uniform distribution with diminishing upper and lower boundaries in each iteration. Those values yielded the best results in our experiments. These percentages and the maximum number of iterations (which is five in our case) also depend on the specific use case and scenario. The central idea is that in early iterations, higher amounts of loads are shifted while the amount decreases with increasing iteration counter. Using random values ensures diversity in the temporary archive. 

\begin{figure}[htb]
    \centering
    \begin{subfigure}[h]{0.49\textwidth}
        \centering
        \includegraphics[width=\textwidth]{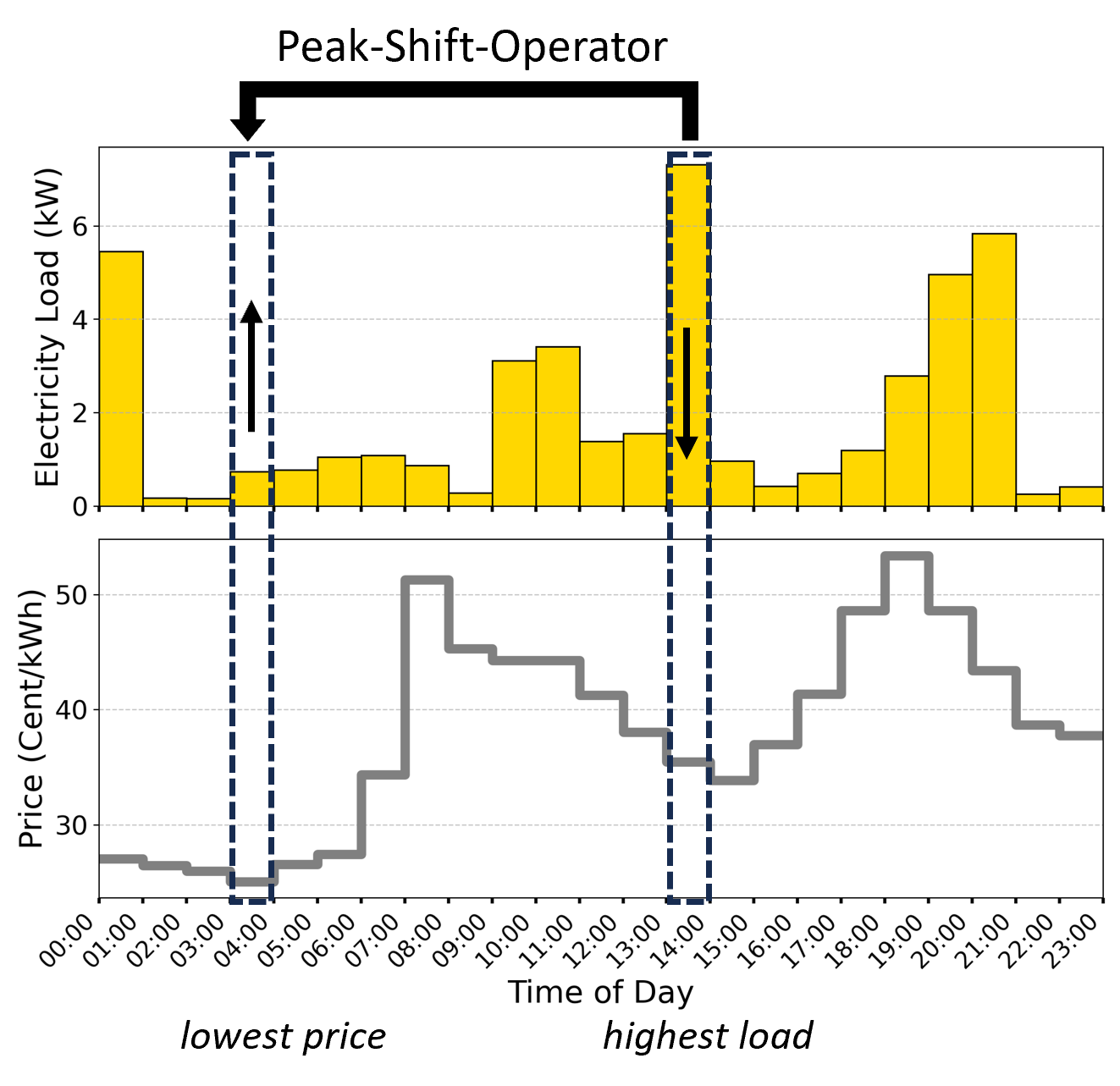}

    \end{subfigure}
   \caption{Peak-Shift-Operator}
    \label{fig_Peak_Shift_Operator}
\end{figure}

The \textit{Peak-Shift-Operator} is similar to the \textit{Price-Shift-Operator} as Fig.~\ref{fig_Peak_Shift_Operator} illustrates. It always chooses the time slot with the highest load in the whole residential area and shifts a certain amount to time slots with low prices. The probabilities for selecting the time slots with the lowest prices are equivalent to the ones of the \textit{Price-Shift-Operator} in Table~\ref{tab_selection_probabilities}. The algorithm likewise randomly determines the amount of shifted load $P_{i}^{\text{PeakShift}\%}$ by using a uniform distribution with decreasing boundaries in each iteration $i$ as shown in Eq.~\ref{eq_amount_peak_shifting}.

\begin{equation}\label{eq_amount_peak_shifting}
\Delta P_{i}^{\text{PeakShift}\%} \sim U(10-i, 25-i), \quad i \in \{1, \ldots, 5\}
\end{equation}

The boundaries can be flexible. We determined the specific values by running multiple experiments and choosing the combination that yielded the best results.\ \textit{PALSS} randomly chooses to either apply the \textit{Peak-Shift-Operator} or the \textit{Price-Shift-Operator} to generate one new candidate solution. This is done several times for every solution in the generation to generate a temporary archive with new candidate solutions.

\subsubsection{Pareto sorting} 
\label{subsubsec_Pareto_sorting}
The number of the new candidate solutions in the temporary archive is higher than the population size. Thus, the \textit{Pareto sorting} eliminates several solutions from it. First, the algorithm determines the Pareto-optimal solutions amongst the candidate solutions of the prior generation. Further, all solutions that violate any constraints are eliminated. A score is calculated by using Eq.~\ref{eq_score} for all remaining solutions. 

\begin{equation}\label{eq_score}
\text{Score} = \frac{\mathrm{Cost}}{\mathrm{Cost}_{\mathrm{Conventional}}} + \frac{P^{\mathrm{Peak}}}{P_{\mathrm{Conventional}}^{\mathrm{Peak}}}
\end{equation}

The equation divides the cost of a solution by the cost of the conventional solution. Likewise, the peak power is divided by the peak of the conventional solution. A lower score indicates a better fulfillment of the two objectives as we deal with a minimization problem.

If the number of Pareto-optimal solutions in the temporary archive of generation $i$ is lower than the population size, all Pareto-optimal solutions are transferred to the next generation $i+1$. The remaining positions in generation $i+1$ are filled by successively adding the dominated solutions of the previous generation's $i$ temporal archive  with the lowest Score value. If the number of Pareto-optimal solutions exceeds the population size, the algorithm only adds the solutions with the lowest (best) Score values to the new generation.

\subsection{Reinforcement learning assisted Pareto local search (RELAPALSS)}
\label{subsec_Reinforcement_learning_assisted_PLS}
Instead of heuristically choosing the time slots and the amount of shifted electrical loads within the \textit{Peak-Shift-Operator} and the \textit{Price-Shift-Operator}, we also investigate the use of model-free reinforcement learning algorithms to learn the parameters for the local search operators. We use the \textit{Proximal Policy Optimization Algorithm} \cite{schulman2017proximal} as it yielded promising results in many different applications and tends to be sample-efficient. We train it on 12 days for one configuration of the residential area with 20 buildings. 

To derive the state for the \textit{Price-Shift-Operator}, the \textit{RELAPALSS} algorithm uses the five time slots with the highest and lowest electricity prices. It then calculates the share of electrical load from the overall load profile of the residential area (as percentages) for each of those time slots. This results in a ten dimensional state space. For the \textit{Peak-Shift-Operator}, the calculation is similar. The difference is that only the five time slots with the lowest electricity prices are considered, making the state space five-dimensional. 

The action space for the \textit{Price-Shift-Operator} is a three-dimensional vector. The first component determines the source time slot for shifting the electricity load. The second component contains the target time slot to shift the load to, while the third quantifies the amount of shifted electrical loads. The action space is merely two-dimensional for the \textit{Peak-Shift-Operator}. The first dimension determines the target time slot to shift the load to, and the second component decides about the amount of shifted electrical loads. 

The \textit{RELAPALSS} algorithm calculates the reward after every application of the local search operators for every building. If the \textit{Price-Shift-Operator} is chosen as the local search operator, the reward quantifies the percentage cost reduction of the individual building multiplied by a constant weighting factor. For the \textit{Peak-Shift-Operator}, the reduced peak load is used as a reward for the \textit{Proximal Policy Optimization Algorithm}.

\subsection{Dichotomous method}
\label{subsec_Dichotomous_method}
\begin{figure}
\begin{center}
\scalebox{0.6}{
  \begin{tikzpicture}
  \pgfplotsset{ticks=none}
      \begin{axis}[grid=none,ymin=-0.5,ymax=10.5,xmax=10,xmin=-0.5,xticklabel=\empty,yticklabel=\empty,minor tick num=0,axis lines = middle,xlabel=$\vn{Cost}$,ylabel=$P^{\vn{Peak}}$,label style =
               {at={(ticklabel cs:1.1)}}]

     \addplot [draw=none,fill=blue!10]coordinates { (1, 10.5) (1, 10) (2, 6) (6, 2.5) (9, 1) (10, 1) (10, 10.5) (1, 10.5)};
      \addplot[mark=*, only marks, color=red,mark size=3pt] coordinates {
        (1, 10)
        (9, 1)
      };
      \addplot[mark=*, only marks, color=red!40,mark size=3pt] coordinates {
        (2, 6)
      };
      \addplot[mark=none, thick, color=red, mark size=3pt] coordinates {
        (1, 10)
        (9, 1)
      };
      \addplot[mark=*, only marks, color=black!20,mark size=3pt] coordinates {
        (6, 2.5)
      };
      \addplot[mark=+, only marks, color=black!20,mark size=3pt] coordinates {
        (1.5, 9)
        (4, 5)
        (5.5, 4)
      };

    \node at (axis cs: 0.4,10) {$P_1$};
    \node at (axis cs: 8.4,0.8) {$P_2$};
    \node at (axis cs: 2,1) {Initial points};
    \end{axis}
  \end{tikzpicture}
  }
  \scalebox{0.6}{
  \begin{tikzpicture}
  \pgfplotsset{ticks=none}
      \begin{axis}[grid=none,ymin=-0.5,ymax=10.5,xmax=10,xmin=-0.5,xticklabel=\empty,yticklabel=\empty,minor tick num=0,axis lines = middle,xlabel=$\vn{Cost}$,ylabel=$P^{\vn{Peak}}$,label style =
               {at={(ticklabel cs:1.1)}}]
      \addplot [draw=none,fill=blue!10]coordinates { (1, 10.5) (1, 10) (2, 6) (6, 2.5) (9, 1) (10, 1) (10, 10.5) (1, 10.5)};
      \addplot[mark=*, only marks, color=red,mark size=3pt] coordinates {
        (1, 10)
        (2, 6)
        (9, 1)
      };
      \addplot[mark=none, thick, color=red, mark size=3pt] coordinates {
        (1, 10)
        (2, 6)
      };
      \addplot[mark=none, thick, color=red, mark size=3pt] coordinates {
        (2, 6)
        (9, 1)
      };
      \addplot[mark=*, only marks, color=red!40,mark size=3pt] coordinates {
        (6, 2.5)
      };
      \addplot[mark=+, only marks, color=black!20,mark size=3pt] coordinates {
        (1.5, 9)
        (4, 5)
        (5.5, 4)
      };

    \node at (axis cs: 0.4,10) {$P_1$};
    \node at (axis cs: 8.4,0.8) {$P_2$};
    \node at (axis cs: 1.4,6) {$P_3$};
    \node at (axis cs: 2,1) {Iteration 1};
    
    \end{axis}
  \end{tikzpicture}
  }
  \scalebox{0.6}{
  \begin{tikzpicture}
  \pgfplotsset{ticks=none}
      \begin{axis}[grid=none,ymin=-0.5,ymax=10.5,xmax=10,xmin=-0.5,xticklabel=\empty,yticklabel=\empty,minor tick num=0,axis lines = middle,xlabel=$C^{total}$,ylabel=$P^{\vn{Peak}}$,label style =
               {at={(ticklabel cs:1.1)}}]
      \addplot [draw=none,fill=blue!10]coordinates { (1, 10.5) (1, 10) (2, 6) (6, 2.5) (9, 1) (10, 1) (10, 10.5) (1, 10.5)};
      \addplot[mark=*, only marks, color=blue,mark size=3pt] coordinates {
        (1, 10)
      };
      \addplot[mark=*, only marks, color=red,mark size=3pt] coordinates {
        (2, 6)
        (6, 2.5)
        (9, 1)
      };
      \addplot[mark=none, thick, color=red, mark size=3pt] coordinates {
        (2, 6)
        (6, 2.5)
      };
      \addplot[mark=none, thick, color=red, mark size=3pt] coordinates {
        (6,2.5)
        (9, 1)
      };
      \addplot[mark=+, only marks, color=black!20,mark size=3pt] coordinates {
        (1.5, 9)
        (4, 5)
        (5.5, 4)
      };

    \node at (axis cs: 0.4,10) {$P_1$};
    \node at (axis cs: 8.4,0.8) {$P_2$};
    \node at (axis cs: 1.4,6) {$P_3$};
    \node at (axis cs: 5.4,2.5) {$P_4$};
    \node at (axis cs: 2,1) {Iteration 2};
    \end{axis}
  \end{tikzpicture}
  }
  \scalebox{0.6}{
  \begin{tikzpicture}
  \pgfplotsset{ticks=none}
      \begin{axis}[grid=none,ymin=-0.5,ymax=10.5,xmax=10,xmin=-0.5,xticklabel=\empty,yticklabel=\empty,minor tick num=0,axis lines = middle,xlabel=$C^{total}$,ylabel=$P^{\vn{Peak}}$,label style =
               {at={(ticklabel cs:1.1)}}]
      \addplot[mark=*, only marks, color=blue,mark size=3pt] coordinates {
        (1, 10)
        (2, 6)
        (6, 2.5)
        (9, 1)
      };
      
      \addplot[mark=+, only marks, color=black,mark size=3pt] coordinates {
        (1.5, 9)
        (4, 5)
        (5.5, 4)
      };
      \addplot [draw=none,fill=blue!10]coordinates { (1, 10.5) (1, 10) (2, 6) (6, 2.5) (9, 1) (10, 1) (10, 10.5) (1, 10.5)};
    
    \node at (axis cs: 3.5,1) {Supported Pareto-front};
    \legend{supported, non-supported}
    \end{axis}
  \end{tikzpicture}
  }
\end{center}
    \caption{Illustration of dichotomous method and (non-)supported Pareto-optimal points.}
    \label{fig:dichotomous}
\end{figure}
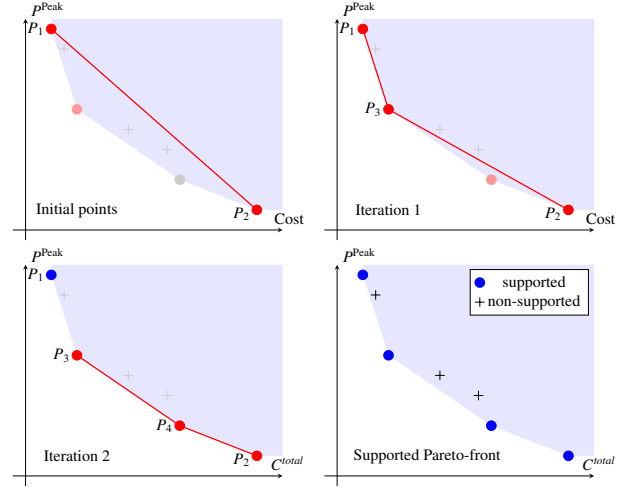

One approach to solving multi-objective optimization problems is scalarization. Here, a weighted sum scalarization is applied to the multi-objective problem with objective functions in Eq.~\ref{eq:objective_function} to obtain a \textit{weighted sum objective}
\begin{equation}\label{eq:weighted_objective_function}
    \min \lambda \vn{Cost} + (1-\lambda) P^{\vn{Peak}}, 
\end{equation}
where $\lambda \in (0,1)$. The multi-objective load shift problem with objective Eq.~\ref{eq:weighted_objective_function} is now a single-objective MILP. For every choice of $\lambda \in (0,1)$ it is guaranteed that a \textit{supported} Pareto-optimal solution is obtained (and all such solutions can be found) \cite{Ehrgott05}. The so-called dichotomous approach is known to be an effective strategy for exploring all necessary weights to find supported Pareto efficient solutions \cite[e.g.][]{Cohon78}. 
Initially the two lexicographically optimal solutions of the problem are computed which prioritise minimising a single objective ($\vn{Cost}$ or $P^{\vn{Peak}}$) and among all optimal solutions of this first problem then minimise the other objective, to obtain the initial two points $P_1, P_2$ on the Pareto-front in Fig.~\ref{fig:dichotomous}. 
The dichotomous method iterates by updating the weight $\lambda$ and solving a weighted sum problem for each pair of neighbouring points until no new points are found -- intuitively the next point is the one furthest from the line connecting the two neighbouring points in Fig.~\ref{fig:dichotomous}. The process terminates with a so-called complete set of extreme supported Pareto-optimal solutions. It should be noted that, unlike a multi-objective continuous linear programme, a multi-objective MILP also has other \textit{non-supported} Pareto-optimal solutions (lying in the interior of the convex hull of supported Pareto-optimal solutions, see bottom-right of Fig.~\ref{fig:dichotomous}) that cannot be obtained by the dichotomous method. Hence, the dichotomous method obtains a subset of the true Pareto-front.

\subsection{Conventional control}
\label{subsec_Conventional_control}

In our application, the conventional control uses a hysteresis-based two-point control approach for the heat pumps for DHW similar to the one described in \cite{BOAIT2012160}. The heat pump starts heating the DHW tank with full power when its usable hot water volume reaches a lower limit. After the volume reaches an upper limit, the heating stops.

For space heating, the heat pump's modulation degree is continuously adjusted to have a constant temperature if possible. Otherwise, the conventional control also uses a small hysteresis to control the deviations from the desired temperature. This control approach is typically used for inverter heat pumps, which can not only be switched on and off but are also capable of modulating their power output \cite{GOMES2013110}. 

Further, the conventional control approach uses uncontrolled charging \cite{Qian2011} for the EV and thus charges the EV with full power if it is connected to the charging station until the battery is fully charged.

%% file: sections/05_results.tex
\section{Results}
\label{sec:Results}

\subsection{Scenarios for the analysis}
\label{sec:Scenarios for the analysis}

Table~\ref{tab_paramters_for_the_simulation} lists the relevant parameters for the residential area and the corresponding sources for them. We investigate three configurations of the residential area: one with ten buildings (4~BT1, 4~BT2, 2~BT3), one with 20 buildings (8~BT1, 8~BT2, 4~BT3), and one with 30 buildings (12~BT1, 12~BT2, 6~BT3). For the electricity price, we use the hourly data of the Day-Ahead market in Germany from 2021, which we took from the \textit{ENTSO-E Transparency Platform} \cite{entsoe_2023}. We scaled the prices up so that they are in line with the average electricity price for residential customers in Germany. We use the software tool \textit{DHWcalc} \cite{jordan2005dhwcalc} to generate the DHW demand profiles. The time resolution $\Delta t$ for the optimization and the simulations is 30 minutes and the optimization horizon is one day.

\begin{table*} 
    \centering
    \caption{Parameters of the residential area}
    \begin{tabular}{l l l l }
        \hline
        \textbf{Parameter} &  \textbf{Value} & \textbf{Source} & \textbf{Comment}\\
        \hline
        Heated area of the buildings BT1, BT2 &  $140~m^2$ & \cite{schaebisch_hall_2024}  & Average German single family house\\
        Heated area of the building BT3 &  $12 \cdot 75~m^2$ & \cite{ZimmerMeisterHaus_2024}  & One example from a residential area in Germany \\
        Concrete width (underﬂoor heating system) &  $7~cm$ & \cite{German_Institute_For_Standardization_2022}  & DIN standard 18560 for screeds in buildings \\
        Density of concrete &  $2400~\frac{kg}{m^3}$ & \cite{DENGIZ2020116651}  & European standards for concrete EN 206-1 \\
        Heat capacity of concrete &  $1000~\frac{J}{kg\cdot K}$ & \cite{DENGIZ2020116651}  & European standards for concrete EN 206-1 \\
        Supply temperature of the heating system &  $30~^\circ C$ & \cite{DENGIZ2020116651}  & \\
        Temperature range for space heating &  $21 - 23~^\circ C$ &  & Assumption for optimal comfort \\
        DHW tank volume &  $200~l$ & \cite{oekoloco_2024} & For 4 inhabitants \\
        Losses of space heating &  $45~W, 220~W$ &  & Assumption $45~W$ (BT1, BT2),  $220~W$ (BT3)  \\
        Losses of DHW tank &  $35~W$&  \cite{umweltbundesamt_2013} &  2nd highest efﬁciency class (EU regulation 814/2013) \\
        Electrical power of the heat pump BT1, BT2 &  $3~kW$&  \cite{bwp_2024} &  Sized to standard outside temperature in Germany\\
        Electrical power of the heat pump BT3 &  $15~kW $&  \cite{bwp_2024} &  Sized to standard outside temperature in Germany\\
        Minimal modulation of the heat pump &  $20~\%$&  \cite{bosch_compress6800} &  Model: Bosch Compress 6800i AW \\
        COP of air-source heat pump for $\Delta T =28~^\circ C$ &  $4.8$&  \cite{bosch_compress6800} &  Model: Bosch Compress 6800i AW \\
        COP of air-source heat pump for $\Delta T =33~^\circ C$ &  $3.9$&  \cite{bosch_compress6800} &  Model: Bosch Compress 6800i AW \\
        Battery capacity of the EV &  $60~kWh$&  \cite{adac_2018} &  Model: Opel Ampera-e \\
        Charging efficiency of the EV &  $89~\%$&  \cite{DENGIZ2020116651} &  Model: Opel Ampera-e \\
        Maximal charging power for home charging &  $4.6~kW$&  \cite{adac_2024} &  Wallbox: KEBA KeContact P30 \\
        Average length of rides for the EV &  $45~km$&  \cite{MID_2018} &  Assumptions inspired by the German Mobility Study \\

    \end{tabular}

    \label{tab_paramters_for_the_simulation}
\end{table*}

We further use the CREST model \cite{RICHARDSON20101878} in combination with German time use data \cite{destatis_2024} to simulate occupancy behavior which builds the basis for the consistent simulation of electricity and space heating demand. By considering local weather conditions \cite{copernicus_2017}, occupancy profiles, and information on household device equipment \cite{BDEW_2016}, electrical appliance demand profiles are generated. The calculation of the space heating demand is based on a 5R1C model following DIN EN ISO 13790, incorporating internal gains derived from the household simulation and thermal building parameters taken from the TABULA residential building typology \cite{IWU_2015}.

The buildings are all located in the federal state of \textit{Schleswig-Holstein} in the north of Germany. We use data collected in the project \textit{iZEUS} \cite{Paetz2013} for the EV driving patterns and their availability.

\subsection{Evaluation}
\label{sec_Evaluation}
We compare our developed algorithms \textit{PALSS} and \textit{RELAPALSS} to the well-known multi-objective evolutionary algorithms \textit{Non-dominated Sorting Genetic Algorithm II (NSGA-II)} \cite{Deb2002} and \textit{Strength Pareto Evolutionary Algorithm (SPEA-II)} \cite{zitzler2001spea2}. We restrict the runtime for each algorithm to solve the optimization problem for one day to 10 minutes. For the evaluation, we randomly chose 18 days from the heating period in Germany (October to March). 

\begin{figure}[htb]
    \centering
    \begin{subfigure}[h]{0.49\textwidth}
        \centering
        \includegraphics[width=\textwidth]{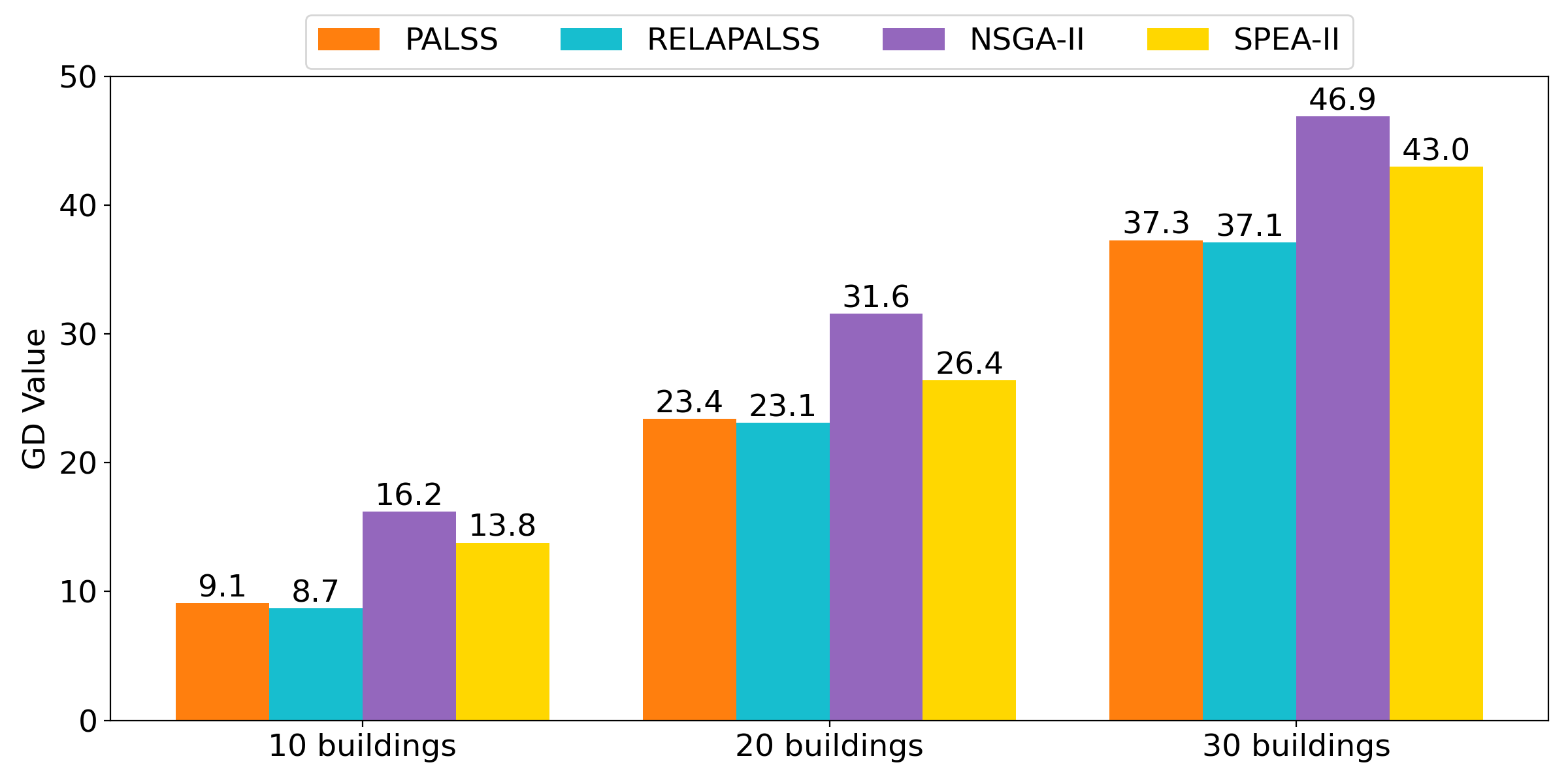}

    \end{subfigure}
   \caption{Results for the Generational Distance (GD) averaged over 18 days and 5 runs}
    \label{fig_results_GD}
\end{figure}

Fig.~\ref{fig_results_GD} shows the results of our analysis for the Generational Distance (GD) averaged over 18 days and 5 runs per method. Thus, each bar represents the average value of 90 runs. GD evaluates the quality of a set of solutions to approximate the true Pareto-front. It quantifies the average distance between the solutions and the points on the true Pareto-front \cite{Ishibuchi2015}. A smaller GD value indicates a better approximation. For all three configurations of the residential area, \textit{PALSS} and \textit{RELAPALSS} yield significantly better results compared to the two metaheuristics \textit{NSGA-II} and \textit{SPEA-II}. Using RL for selecting the relevant parameters of \textit{PALSS} leads to a small additional improvement as in all categories \textit{RELAPALSS} leads to lower GD values on average. \textit{SPEA-II} yields better approximations compared to the other evolutionary algorithm \textit{NSGA-II}. Generally, the GD values increase with an increasing number of residential buildings. This is because the problem complexity strongly increases with a higher number of buildings, and we execute every algorithm for only ten minutes independently of the problem size. Thus, only a smaller fraction of the search space can be explored using our developed approaches and the metaheuristics. All methods lead to crucial improvements compared to the conventional control approach, which, on average, achieved GD values of 22.0 for 10 buildings, 40.4 for 20 buildings, and 60.0 for 30 buildings.

\begin{figure}[htb]
    \centering
    \begin{subfigure}[h]{0.49\textwidth}
        \centering
        \includegraphics[width=\textwidth]{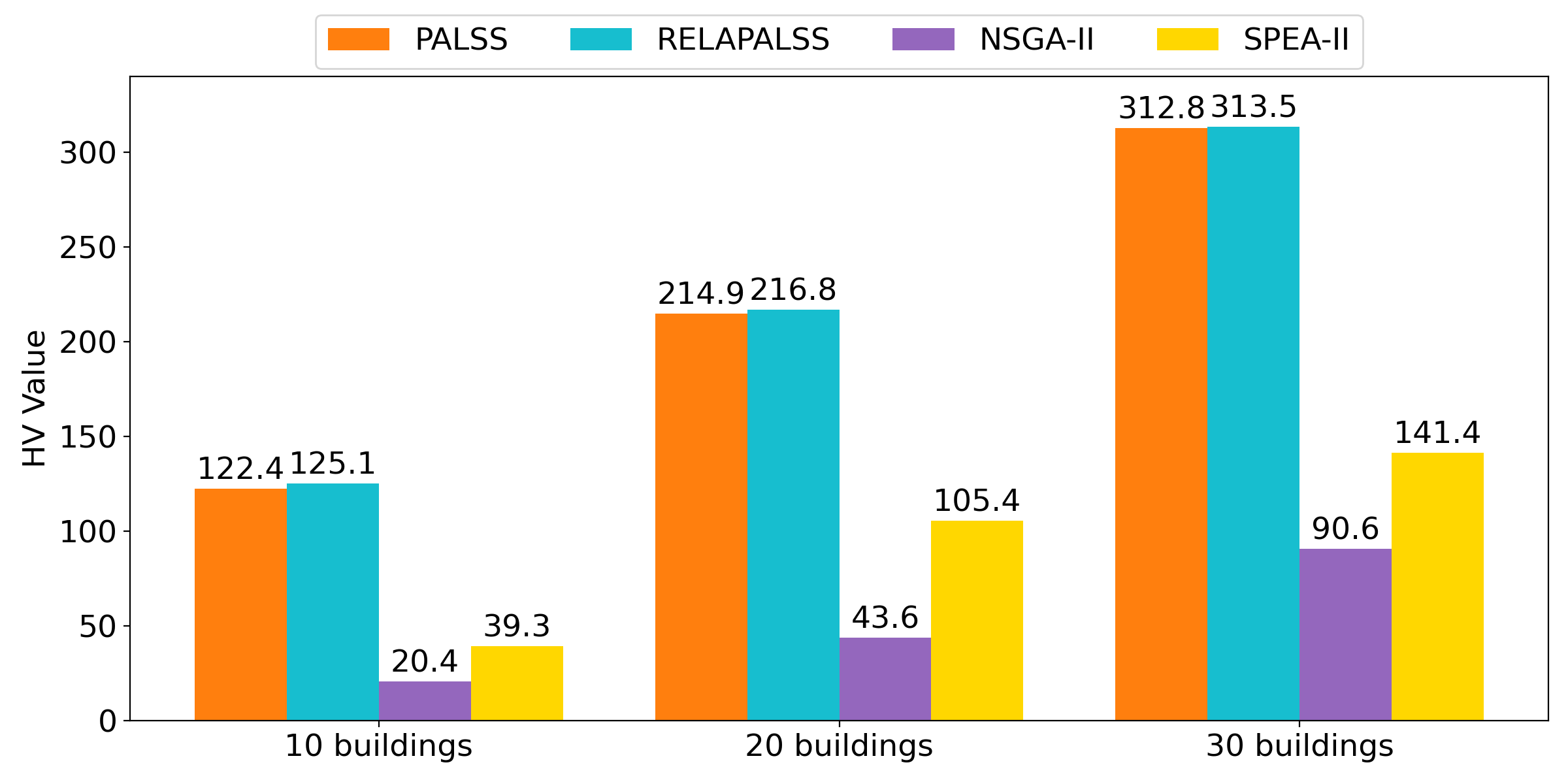}

    \end{subfigure}
   \caption{Results for the Hypervolume (HV) averaged over 18 days and 5 runs}
    \label{fig_results_HV}
\end{figure}

To evaluate the coverage of the objective space by the approximated Pareto-front, we use the hypervolume (HV) indicator \cite{Bader2011}. HV quantifies the hypervolume between an approximated Pareto-front and a reference point. Higher values indicate a better coverage. We choose the conventional control's resulting costs and peak load as the reference point in the two-dimensional objective space. Fig.~\ref{fig_results_HV} depicts the results for HV of our experiments averaged over 18 days and 5 runs. Analogous to GD, the results for all configurations of the residential area reveal that our developed methods \textit{PALSS} and \textit{RELAPALSS} strongly outperform the two metaheuristics \textit{NSGA-II} and \textit{SPEA-II}. \textit{RELAPALSS} similarly leads to slightly better results compared to \textit{PALSS}. In this case, the HV values increase with increasing number of buildings. A larger problem size increases the improvements over the naively generated conventional solution. Thus, based on the reference point, the approximated Pareto-fronts enclose a larger volume. However, the distance to the real Pareto-front becomes larger with higher problem complexity. The average HV values for the dichotomous method are 303.9 for 10 buildings, 933.7 for 20 buildings, and 1687.2 for 30 buildings.

\begin{figure*}[t]
    \centering
    \begin{subfigure}[h]{1.00\textwidth}
        \centering
        \includegraphics[width=\textwidth]{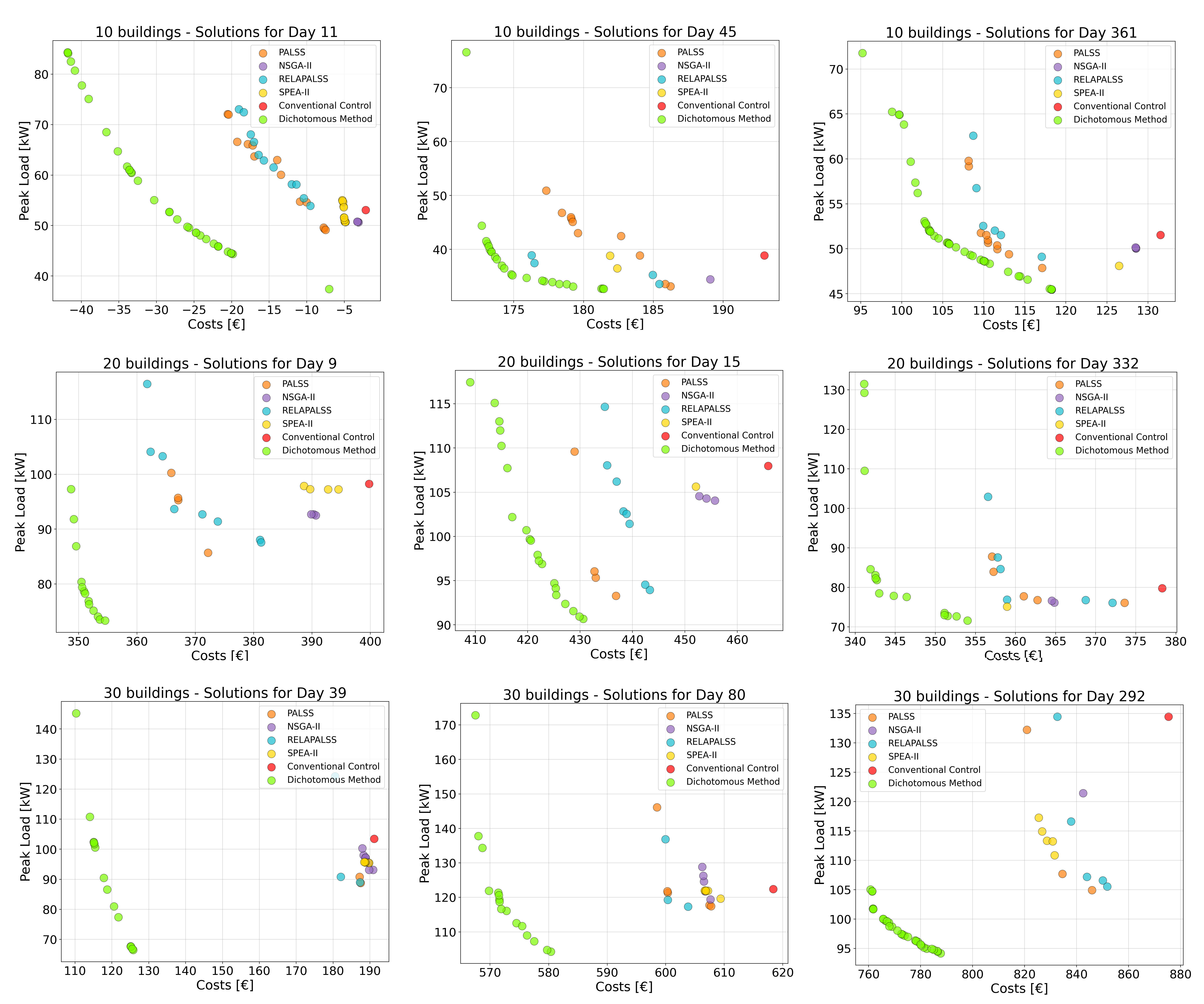}

    \end{subfigure}
   \caption{Pareto fronts of the different methods}
    \label{fig_pareto_fronts_combined}
\end{figure*}

Fig.~\ref{fig_pareto_fronts_combined} illustrates the found Pareto-fronts of all six methods for different days and configurations of the residential area. It can be seen that the distance between the approximated Pareto-fronts of the various methods and the true Pareto-front, obtained by using the dichotomous method, becomes more significant with increasing problem size. As the wholesale prices on the electricity market can also be negative, meaning that you get money when consuming electricity, the resulting costs of day 11 are also negative. The Pareto-fronts for the different days also reveal that cost minimization and peak load minimization are conflicting objectives since minimizing one goal without worsening the other is impossible. 

We tested different parameters for each method and used the ones that yielded the best results. For all four methods \textit{PALSS}, \textit{RELAPALSS}, \textit{NSGA-II}, and \textit{SPEA-II}, the population size was 20. The number of new solutions for each member in the population (offsprings) was 3 for the methods \textit{PALSS} and \textit{RELAPALSS} while for the two metaheuristics, the number was 10. Both \textit{NSGA-II} and \textit{SPEA-II} used the Simulated Binary Crossover (SBX) and started from the solution of the conventional control approach. While the runtime for these four methods was chosen to be 10 minutes, the execution time of the dichotomous method depends on the problem size. For a residential area with 10 buildings, the dichotomous method needed, on average, 42 minutes per day for 20 buildings, 94 minutes, and 122 minutes for a problem size of 30. We use the Gurobi solver to solve the resulting single-objective optimization problems of the dichotomous method \cite{gurobi}. All experiments were carried out on an \textit{Intel Xeon E-1650 v2, 3,5GHz, 6 Cores}. We set the optimality gap for exactly solving the resulting single-objective optimization problems to $0.1\%$ and the maximum runtime for a single problem to 10 minutes.

On average, the training time for \textit{RELAPALSS} was 4 hours, and we trained it on 12 different days, using the conventional solution as the starting solution. It should be noted that we trained the RL model solely on a residential area consisting of 20 buildings. We subsequently applied this trained model to other configurations of the residential area, such as those with 10 and 30 buildings, without retraining. The results reveal that the trained model is thus scalable and can be applied to different sizes of the residential area.  

We implemented the simulations and the optimization in the programming language \textit{Python}. For the MILP we use the package \textit{Pyomo} \cite{hart2011pyomo}, for the evolutionary algorithms \textit{Pymoo} \cite{Pymoo}, for the RL \textit{Stable-Baselines3} \textit{stable-baselines3} \cite{stable-baselines3} and \textit{gymnasium} \cite{towers_gymnasium_2023}. All methods use a super-ordinate additional controller that adjusts their actions if a constraint violation is about to occur. 

Fig.~\ref{fig_Profiles_combined} in the appendix plots exemplary profiles for one day (Day 332) resulting from the different control approaches and input profiles (DHW, Space heating, number of EVs, outside temperature, and price). The profiles result from a residential area with 20 buildings (8 BT1, 8 BT2, 4 BT3).

\subsection{Critical appraisal}
\label{sec_Critical appraisal}
We made some simplifications in our study to carry out the experiments. To solve the optimization problem optimally, we assumed a perfect forecast of future demands. Thus, we did not incorporate any uncertainties. To cope with uncertainties of energy consumption, forecasts of energy demand, for example, using artificial neural networks, are necessary \cite{NETO20082169}. Further, the building model for the optimization assumes one uniform temperature in the whole building and does not consider heat flows between different rooms or internal heat gains. Moreover, we use a centralized control approach to control the flexible loads. While this typically leads to the best results, it might infringe on the privacy of the inhabitants \cite{DENGIZ2021119984}. 
It has to be noted that our introduced control mechanisms do not require any forecast and are model-free approaches that do not rely on any building model. Further, they can be combined with decentralized scheduling-based control approaches like the ones introduced in the study \cite{DENGIZ2021119984}. Hence, we are convinced that our approaches can be used in many simulations and real-world applications. 
We also set a maximum runtime for the single-objective optimization problems that the dichotomous method repeatedly solves. Thus, it could happen that the exact methods did not solve the optimization problem to optimality. Because of this, what we refer to as the true Pareto-front that we use for evaluation, is itself only an approximation. However, this approximation is very close to the theoretical true front. 

%% file: sections/06_conclusion.tex
\section{Conclusion}
\label{sec:Conclusion}
In this paper, we introduce two novel approaches for the multi-objective optimization of the conflicting goals to minimize the electricity cost and the peak load in a residential area. The approaches intelligently control flexible electrical devices like heat pumps and electric vehicles such that the electricity consumption of the residential area reacts to an external electricity price signal. The two approaches  \textit{Pareto local search for shifting loads (PALSS)} and \textit{Reinforcement learning assisted Pareto local for shifting loads (RELAPALSS)} define a concept for Pareto local search with heuristic neighborhood searching operations. \textit{RELAPALSS} additionally uses reinforcement learning to adjust the parameters of the local search operations. 

Our experiments on a simulated residential area with different building types and varying numbers of buildings show that our developed methods outperform the two state-of-the-art evolutionary algorithms \textit{Non-dominated Sorting Genetic Algorithm II (NSGA-II)} and \textit{Strength Pareto Evolutionary Algorithm (SPEA-II)} regarding the two relevant evaluation metrics for multi-objective optimization \textit{Generational Distance (GD)} and \textit{Hypervolume (HV)}. Including reinforcement learning in the local search operation leads to additional small improvements. Our developed methods do not depend on any model or a forecast and can be applied to other flexible electrical appliances. They significantly reduce costs and peak loads compared to conventional control approaches and metaheuristics. Thus, they can contribute to better utilizing volatile renewable energy sources in residential areas. 

In future work, we intend to analyze the use of imitation learning in combination with domain knowledge to improve our local search approaches further. Moreover, we plan to integrate other flexibility options like batteries or additional electric heating elements in gas heating systems into our model. Also, combining our developed methods with decentralized coordination approaches constitutes a possible future task.

%% file: sections/ending.tex
\section*{Supplementary materials}
\label{sec_supplementary_materials}
Both the code and data are openly accessible. The repository containing the commented code and an executable notebook is available at 
\href{https://github.com/thomasdengiz/MultiObjective_LocalSearch_DSM}{GitHub}. 
The input data and result profiles can be accessed at 
\href{https://publikationen.bibliothek.kit.edu/1000170063}{KITOpen}.

\section*{CRediT author statement}
\label{sec_Credit_author_statement}
\textbf{Thomas Dengiz}: Conceptualization, Methodology, Software, Validation, Formal analysis, Investigation, Writing - Original Draft, Visualization. \textbf{Andrea Raith}: Conceptualization, Methodology, Investigation, Writing - Original Draft. \textbf{Max Kleinebrahm}: Validation, Resources, Data Curation, Writing - Original Draft. \textbf{Jonathan Vogl}: Validation, Software, Writing - Review \& Editing.
\textbf{Wolf Fichtner}: Validation, Writing - Review \& Editing, Funding acquisition.

\section*{Acknowledgments}
\label{sec_acknowledgments}
This work was supported by the project AsimutE (Autoconsommation et Stockage Intelligents pour une Meilleure Utilisation de l’Énergie) from the European Territorial Cooperation program Interreg. The authors also want to thank the Karlsruhe House of Young Scientists (KHYS) for enabling a research stay, which made the necessary collaborations for this paper possible.   

\section*{Appendix}
\label{sec:appendix}
 \begin{figure*}[t]
    \centering
    \begin{subfigure}[h]{1.0\textwidth}
        \centering
        \includegraphics[width=\textwidth]{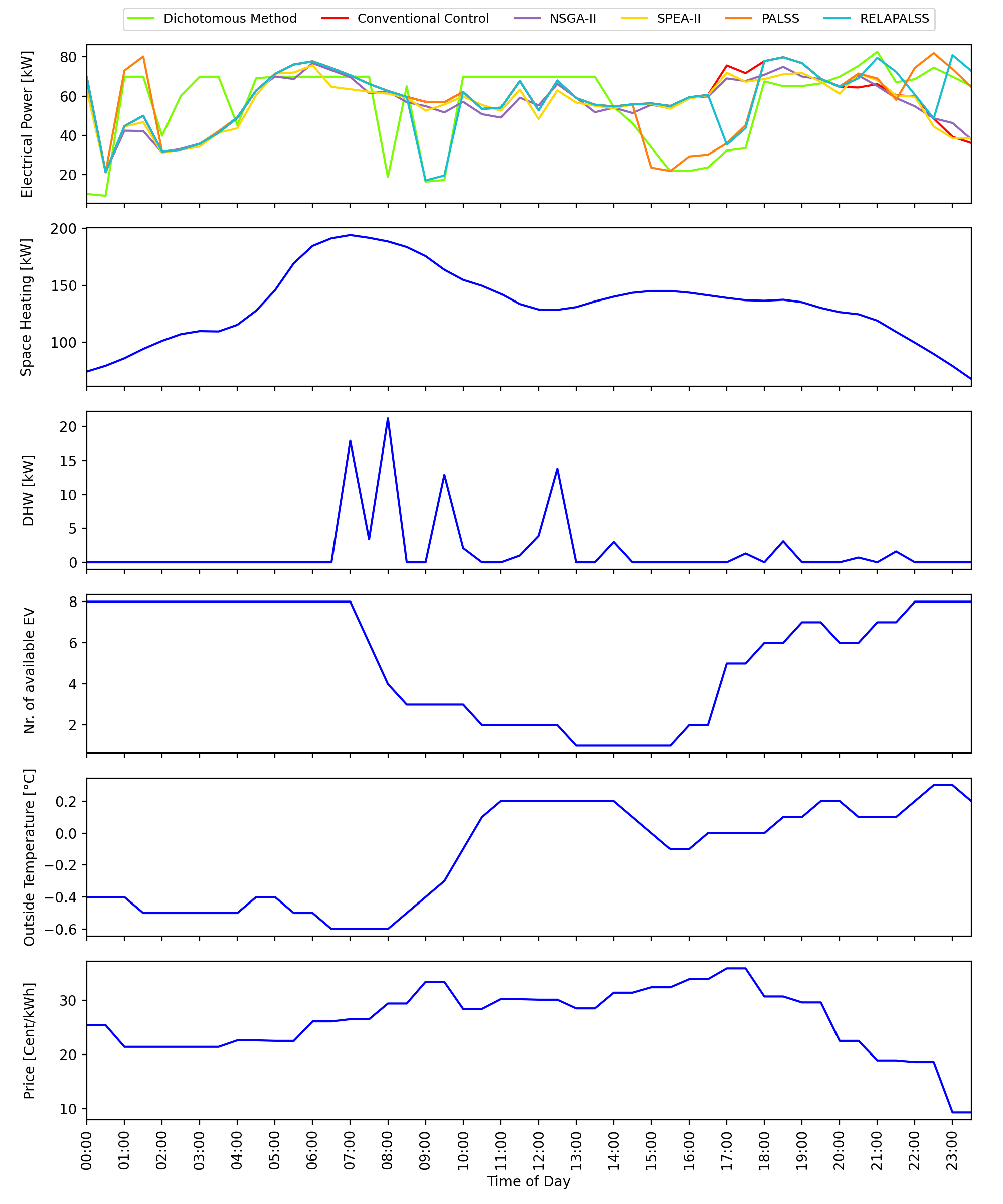}

    \end{subfigure}
   \caption{Exemplary profiles for one day in November for a residential area with 20 buildings}
    \label{fig_Profiles_combined}
\end{figure*}

Exemplary profiles for one day in November for a residential area with 20 buildings are presented in Fig.~\ref{fig_Profiles_combined}.

%% file: main.bbl
\begin{thebibliography}{10}
\expandafter\ifx\csname url\endcsname\relax
  \def\url#1{\texttt{#1}}\fi
\expandafter\ifx\csname urlprefix\endcsname\relax\def\urlprefix{URL }\fi
\expandafter\ifx\csname href\endcsname\relax
  \def\href#1#2{#2} \def\path#1{#1}\fi

\bibitem{Shao2011}
S.~Shao, M.~Pipattanasomporn, S.~Rahman, Demand response as a load shaping tool in an intelligent grid with electric vehicles, IEEE Transactions on Smart Grid 2~(4) (2011) 624--631.
\newblock \href {https://doi.org/https://doi.org/10.1109/TSG.2011.2164583} {\path{doi:https://doi.org/10.1109/TSG.2011.2164583}}.

\bibitem{Patteeuw2015}
D.~Patteeuw, K.~Bruninx, A.~Arteconi, E.~Delarue, W.~D’haeseleer, L.~Helsen, Integrated modeling of active demand response with electric heating systems coupled to thermal energy storage systems, Applied Energy 151 (2015) 306--319.
\newblock \href {https://doi.org/https://doi.org/10.1016/j.apenergy.2015.04.014} {\path{doi:https://doi.org/10.1016/j.apenergy.2015.04.014}}.

\bibitem{SHARIATZADEH2015}
F.~Shariatzadeh, P.~Mandal, A.~K. Srivastava, Demand response for sustainable energy systems: A review, application and implementation strategy, Renewable and Sustainable Energy Reviews 45 (2015) 343--350.
\newblock \href {https://doi.org/https://doi.org/10.1016/j.rser.2015.01.062} {\path{doi:https://doi.org/10.1016/j.rser.2015.01.062}}.

\bibitem{KUHNBACH2021}
M.~Kühnbach, J.~Stute, A.-L. Klingler, Impacts of avalanche effects of price-optimized electric vehicle charging - does demand response make it worse?, Energy Strategy Reviews 34 (2021) 100608.
\newblock \href {https://doi.org/https://doi.org/10.1016/j.esr.2020.100608} {\path{doi:https://doi.org/10.1016/j.esr.2020.100608}}.

\bibitem{ULLMAN1975}
J.~Ullman, Np-complete scheduling problems, Journal of Computer and System Sciences 10~(3) (1975) 384--393.
\newblock \href {https://doi.org/https://doi.org/10.1016/S0022-0000(75)80008-0} {\path{doi:https://doi.org/10.1016/S0022-0000(75)80008-0}}.

\bibitem{Luo2022}
Q.~Luo, G.~Wu, B.~Ji, L.~Wang, P.~N. Suganthan, Hybrid multi-objective optimization approach with pareto local search for collaborative truck-drone routing problems considering flexible time windows, IEEE Transactions on Intelligent Transportation Systems 23~(8) (2022) 13011--13025.
\newblock \href {https://doi.org/https://doi.org/10.1109/TITS.2021.3119080} {\path{doi:https://doi.org/10.1109/TITS.2021.3119080}}.

\bibitem{Paquete2004}
L.~Paquete, M.~Chiarandini, T.~St{\"u}tzle, Pareto local optimum sets in the biobjective traveling salesman problem: An experimental study, in: X.~Gandibleux, M.~Sevaux, K.~S{\"o}rensen, V.~T'kindt (Eds.), Metaheuristics for Multiobjective Optimisation, Springer Berlin Heidelberg, Berlin, Heidelberg, 2004, pp. 177--199.
\newblock \href {https://doi.org/https://doi.org/10.1007/978-3-642-17144-4_7} {\path{doi:https://doi.org/10.1007/978-3-642-17144-4_7}}.

\bibitem{PINTO20211}
G.~Pinto, M.~S. Piscitelli, J.~R. Vázquez-Canteli, Z.~Nagy, A.~Capozzoli, Coordinated energy management for a cluster of buildings through deep reinforcement learning, Energy 229 (2021) 120725.
\newblock \href {https://doi.org/https://doi.org/10.1016/j.energy.2021.120725} {\path{doi:https://doi.org/10.1016/j.energy.2021.120725}}.

\bibitem{PEIRELINCK2024}
T.~Peirelinck, C.~Hermans, F.~Spiessens, G.~Deconinck, Combined peak reduction and self-consumption using proximal policy optimisation, Energy and AI 16 (2024) 100323.
\newblock \href {https://doi.org/https://doi.org/10.1016/j.egyai.2023.100323} {\path{doi:https://doi.org/10.1016/j.egyai.2023.100323}}.

\bibitem{PINTO20211_2}
G.~Pinto, D.~Deltetto, A.~Capozzoli, Data-driven district energy management with surrogate models and deep reinforcement learning, Applied Energy 304 (2021) 117642.
\newblock \href {https://doi.org/https://doi.org/10.1016/j.apenergy.2021.117642} {\path{doi:https://doi.org/10.1016/j.apenergy.2021.117642}}.

\bibitem{SHAKOURIG2017}
H.~{Shakouri G.}, A.~Kazemi, Multi-objective cost-load optimization for demand side management of a residential area in smart grids, Sustainable Cities and Society 32 (2017) 171--180.
\newblock \href {https://doi.org/https://doi.org/10.1016/j.scs.2017.03.018} {\path{doi:https://doi.org/10.1016/j.scs.2017.03.018}}.

\bibitem{Tang2022}
H.~Tang, Y.~Li, D.~Li, H.~Wang, Dynamic optimization of on-grid integrated energy system considering peak-shaving demand via learning methods, IEEJ Transactions on Electrical and Electronic Engineering 17~(10) (2022) 1409--1419.
\newblock \href {https://doi.org/https://doi.org/10.1002/tee.23651} {\path{doi:https://doi.org/10.1002/tee.23651}}.

\bibitem{Dutra2019}
M.~D. de~Souza~Dutra, M.~F. Anjos, S.~L. Digabel, A framework for peak shaving through the coordination of smart homes, in: 2019 IEEE PES Innovative Smart Grid Technologies Conference - Latin America (ISGT Latin America), 2019, pp. 1--6.
\newblock \href {https://doi.org/https://doi.org/10.1109/ISGT-LA.2019.8895476} {\path{doi:https://doi.org/10.1109/ISGT-LA.2019.8895476}}.

\bibitem{Khalid2018}
A.~Khalid, N.~Javaid, M.~Guizani, M.~Alhussein, K.~Aurangzeb, M.~Ilahi, Towards dynamic coordination among home appliances using multi-objective energy optimization for demand side management in smart buildings, IEEE Access 6 (2018) 19509--19529.
\newblock \href {https://doi.org/https://doi.org/10.1109/ACCESS.2018.2791546} {\path{doi:https://doi.org/10.1109/ACCESS.2018.2791546}}.

\bibitem{Wynn2022}
S.~L.~L. Wynn, W.~Pinthurat, B.~Marungsri, Multi-objective optimization for peak shaving with demand response under renewable generation uncertainty, Energies 15~(23) (2022).
\newblock \href {https://doi.org/https://doi.org/10.3390/en15238989} {\path{doi:https://doi.org/10.3390/en15238989}}.

\bibitem{Terlouw2019}
T.~Terlouw, T.~AlSkaif, C.~Bauer, W.~{van Sark}, Multi-objective optimization of energy arbitrage in community energy storage systems using different battery technologies, Applied Energy 239 (2019) 356--372.
\newblock \href {https://doi.org/https://doi.org/10.1016/j.apenergy.2019.01.227} {\path{doi:https://doi.org/10.1016/j.apenergy.2019.01.227}}.

\bibitem{Song2022}
Z.~Song, X.~Guan, M.~Cheng, Multi-objective optimization strategy for home energy management system including pv and battery energy storage, Energy Reports 8 (2022) 5396--5411.
\newblock \href {https://doi.org/https://doi.org/10.1016/j.egyr.2022.04.023} {\path{doi:https://doi.org/10.1016/j.egyr.2022.04.023}}.

\bibitem{Wang2020}
J.~Wang, Y.~Liu, F.~Ren, S.~Lu, Multi-objective optimization and selection of hybrid combined cooling, heating and power systems considering operational flexibility, Energy 197 (2020) 117313.
\newblock \href {https://doi.org/https://doi.org/10.1016/j.energy.2020.117313} {\path{doi:https://doi.org/10.1016/j.energy.2020.117313}}.

\bibitem{Kazmi2019}
H.~Kazmi, J.~Suykens, A.~Balint, J.~Driesen, Multi-agent reinforcement learning for modeling and control of thermostatically controlled loads, Applied Energy 238 (2019) 1022--1035.
\newblock \href {https://doi.org/https://doi.org/10.1016/j.apenergy.2019.01.140} {\path{doi:https://doi.org/10.1016/j.apenergy.2019.01.140}}.

\bibitem{Wei2015}
X.~Wei, A.~Kusiak, M.~Li, F.~Tang, Y.~Zeng, Multi-objective optimization of the hvac (heating, ventilation, and air conditioning) system performance, Energy 83 (2015) 294--306.
\newblock \href {https://doi.org/https://doi.org/10.1016/j.energy.2015.02.024} {\path{doi:https://doi.org/10.1016/j.energy.2015.02.024}}.

\bibitem{Chiu2020}
W.-Y. Chiu, J.-T. Hsieh, C.-M. Chen, Pareto optimal demand response based on energy costs and load factor in smart grid, IEEE Transactions on Industrial Informatics 16~(3) (2020) 1811--1822.
\newblock \href {https://doi.org/https://doi.org/10.1016/10.1109/TII.2019.2928520} {\path{doi:https://doi.org/10.1016/10.1109/TII.2019.2928520}}.

\bibitem{schulman2017proximal}
J.~Schulman, F.~Wolski, P.~Dhariwal, A.~Radford, O.~Klimov, Proximal policy optimization algorithms (2017).
\newblock \href {http://arxiv.org/abs/1707.06347} {\path{arXiv:1707.06347}}, \href {https://doi.org/https://doi.org/10.48550/arXiv.1707.06347} {\path{doi:https://doi.org/10.48550/arXiv.1707.06347}}.

\bibitem{Ehrgott05}
M.~Ehrgott, Multicriteria Optimization, second edition Edition, Springer, 2005.

\bibitem{Cohon78}
J.~Cohon, Multiobjective Programming and Planning, Academic Press, New York, 1978.

\bibitem{BOAIT2012160}
P.~Boait, D.~Dixon, D.~Fan, A.~Stafford, Production efficiency of hot water for domestic use, Energy and Buildings 54 (2012) 160--168.
\newblock \href {https://doi.org/https://doi.org/10.1016/j.enbuild.2012.07.011} {\path{doi:https://doi.org/10.1016/j.enbuild.2012.07.011}}.

\bibitem{GOMES2013110}
A.~Gomes, C.~H. Antunes, J.~Martinho, A physically-based model for simulating inverter type air conditioners/heat pumps, Energy 50 (2013) 110--119.
\newblock \href {https://doi.org/https://doi.org/10.1016/j.energy.2012.11.047} {\path{doi:https://doi.org/10.1016/j.energy.2012.11.047}}.

\bibitem{Qian2011}
K.~Qian, C.~Zhou, M.~Allan, Y.~Yuan, Modeling of load demand due to ev battery charging in distribution systems, IEEE Transactions on Power Systems 26~(2) (2011) 802--810.
\newblock \href {https://doi.org/https://doi.org/10.1109/TPWRS.2010.2057456} {\path{doi:https://doi.org/10.1109/TPWRS.2010.2057456}}.

\bibitem{entsoe_2023}
{Entso-E Transparency Platform}, \href{https://transparency.entsoe.eu/}{Day-ahead prices}, accessed: 24.04.2024 (2023).
\newline\urlprefix\url{https://transparency.entsoe.eu/}

\bibitem{jordan2005dhwcalc}
U.~Jordan, K.~Vajen, \href{https://www.uni-kassel.de/maschinenbau/en/institute/thermische-energietechnik/fachgebiete/solar-und-anlagentechnik/downloads}{Dhwcalc: Program to generate domestic hot water profiles with statistical means for user defined conditions}, accessed: 24.04.2024 (2005).
\newline\urlprefix\url{https://www.uni-kassel.de/maschinenbau/en/institute/thermische-energietechnik/fachgebiete/solar-und-anlagentechnik/downloads}

\bibitem{schaebisch_hall_2024}
{Bausparkasse Schwäbisch Hall AG}, \href{https://www.schwaebisch-hall.de/ratgeber/neubau-und-anbau/zusammen-leben/das-ideale-familienhaus.html}{Ein {E}infamilienhaus bauen: {S}o wird der {T}raum {W}irklichkeit}, accessed: 24.04.2024 (2024).
\newline\urlprefix\url{https://www.schwaebisch-hall.de/ratgeber/neubau-und-anbau/zusammen-leben/das-ideale-familienhaus.html}

\bibitem{ZimmerMeisterHaus_2024}
{ZimmerMeisterHaus Service \& Dienstleistungs GmbH}, \href{https://www.zmh.com/wohnbaugewerbebau/beispiele-wohn-gewerbebau/portrait/mehrfamilienhaus-oepfingen.html}{Mehrfamilienhaus Öpfingen}, accessed: 24.04.2024 (2024).
\newline\urlprefix\url{https://www.zmh.com/wohnbaugewerbebau/beispiele-wohn-gewerbebau/portrait/mehrfamilienhaus-oepfingen.html}

\bibitem{German_Institute_For_Standardization_2022}
{German Institute for Standardization}, \href{https://www.beuth.de/de/norm/din-18560-2/356233036}{Din 18560-2:2022-08 estriche im bauwesen - teil 2: Estriche und heizestriche auf dämmschichten}, accessed: 24.04.2024 (2022).
\newline\urlprefix\url{https://www.beuth.de/de/norm/din-18560-2/356233036}

\bibitem{DENGIZ2020116651}
T.~Dengiz, P.~Jochem, \href{https://www.sciencedirect.com/science/article/pii/S0360544219323461}{Decentralized optimization approaches for using the load flexibility of electric heating devices}, Energy 193 (2020) 116651.
\newblock \href {https://doi.org/https://doi.org/10.1016/j.energy.2019.116651} {\path{doi:https://doi.org/10.1016/j.energy.2019.116651}}.
\newline\urlprefix\url{https://www.sciencedirect.com/science/article/pii/S0360544219323461}

\bibitem{oekoloco_2024}
{ökoloco GmbH}, \href{https://oekoloco.de/heizungen/heizungskomponenten/warmwasserspeicher/}{Warmwasserspeicher: Vor- und nachteile, größe und hersteller}, accessed: 24.04.2024 (2024).
\newline\urlprefix\url{https://oekoloco.de/heizungen/heizungskomponenten/warmwasserspeicher/}

\bibitem{umweltbundesamt_2013}
{Umweltbundesamt}, \href{https://www.umweltbundesamt.de/sites/default/files/medien/376/dokumente/oekodesignrichtlinie_und_energieverbrauchskennzeichnung_warmwasserbereiter.pdf}{Ökodesign-richtlinie und energieverbrauchskennzeichnung - warmwasserbereiter und warmwasserspeicher}, accessed: 24.04.2024 (2013).
\newline\urlprefix\url{https://www.umweltbundesamt.de/sites/default/files/medien/376/dokumente/oekodesignrichtlinie_und_energieverbrauchskennzeichnung_warmwasserbereiter.pdf}

\bibitem{bwp_2024}
{Bundesverband Wärmepumpe (BWP) e.V.}, \href{https://www.waermepumpe.de/normen-technik/klimakarte/}{Klimakarte}, accessed: 24.04.2024 (2024).
\newline\urlprefix\url{https://www.waermepumpe.de/normen-technik/klimakarte/}

\bibitem{bosch_compress6800}
{Bosch Thermotechnik GmbH}, \href{https://www.bosch-homecomfort.com/de/de/ocs/wohngebaeude/compress-6800i-aw-19312695-p/}{Luft-wasser-wärmepumpe compress 6800i aw}, accessed: 24.04.2024 (2024).
\newline\urlprefix\url{https://www.bosch-homecomfort.com/de/de/ocs/wohngebaeude/compress-6800i-aw-19312695-p/}

\bibitem{adac_2018}
{Allgemeiner Deutscher Automobil-Club e.V. (ADAC)}, \href{https://www.adac.de/rund-ums-fahrzeug/autokatalog/marken-modelle/opel/ampera/2generation/275080/}{Opel ampera-e first edition (07/17 - 01/18)}, accessed: 24.04.2024 (2018).
\newline\urlprefix\url{https://www.adac.de/rund-ums-fahrzeug/autokatalog/marken-modelle/opel/ampera/2generation/275080/}

\bibitem{adac_2024}
{Allgemeiner Deutscher Automobil-Club e.V. (ADAC)}, \href{https://www.adac.de/rund-ums-fahrzeug/elektromobilitaet/tests/wallboxen/keba-kecontact-p30-id-4180/}{Keba kecontact p30}, accessed: 24.04.2024 (2024).
\newline\urlprefix\url{https://www.adac.de/rund-ums-fahrzeug/elektromobilitaet/tests/wallboxen/keba-kecontact-p30-id-4180/}

\bibitem{MID_2018}
{ Federal ministry of transport and digital infrastructure}, \href{https://bmdv.bund.de/SharedDocs/DE/Anlage/G/mid-ergebnisbericht.pdf?__blob=publicationFile}{Mobilität in deutschland - {M}i{D}}, accessed: 24.04.2024 (2018).
\newline\urlprefix\url{https://bmdv.bund.de/SharedDocs/DE/Anlage/G/mid-ergebnisbericht.pdf?__blob=publicationFile}

\bibitem{RICHARDSON20101878}
I.~Richardson, M.~Thomson, D.~Infield, C.~Clifford, Domestic electricity use: A high-resolution energy demand model, Energy and Buildings 42~(10) (2010) 1878--1887.
\newblock \href {https://doi.org/https://doi.org/10.1016/j.enbuild.2010.05.023} {\path{doi:https://doi.org/10.1016/j.enbuild.2010.05.023}}.

\bibitem{destatis_2024}
{Statistisches Bundesamt (Destatis)}, \href{https://www.destatis.de/DE/Themen/Gesellschaft-Umwelt/Einkommen-Konsum-Lebensbedingungen/Zeitverwendung/_inhalt.html#sprg234984}{Einkommen, konsum und lebens bedingungen - zeitverwendung}, accessed: 24.04.2024 (2024).
\newline\urlprefix\url{https://www.destatis.de/DE/Themen/Gesellschaft-Umwelt/Einkommen-Konsum-Lebensbedingungen/Zeitverwendung/_inhalt.html#sprg234984}

\bibitem{copernicus_2017}
{Copernicus Climate Change Service}, \href{https://cds.climate.copernicus.eu/#!/search?text=ERA5&type=dataset}{Era5: Fifth generation of ecmwf atmospheric reanalyses of the global climate - datasets} (2017).
\newline\urlprefix\url{https://cds.climate.copernicus.eu/#!/search?text=ERA5&type=dataset}

\bibitem{BDEW_2016}
{BDEW Bundesverband der Energie- und Wasserwirtschaft e.V.}, \href{https://www. bdew.de/media/documents/BDEWStromverbrauchimHaushaltJuli2016Charts.pdf}{Stromverbrauch im haushalt 2016}, accessed: 17.03.2016 (2016).
\newline\urlprefix\url{https://www. bdew.de/media/documents/BDEWStromverbrauchimHaushaltJuli2016Charts.pdf}

\bibitem{IWU_2015}
{Institut Wohnen und Umwelt GmbH}, \href{https://www.episcope.eu/downloads/public/docs/brochure/DE_TABULA_TypologyBrochure_IWU.pdf}{Deutsche wohngebäudetypologie. beispielhafte maßnahmen zur verbesserung der energieeffizienz von typischen wohngebäuden – zweite erweiterte auflage –}, accessed: 24.04.2024 (2015).
\newline\urlprefix\url{https://www.episcope.eu/downloads/public/docs/brochure/DE_TABULA_TypologyBrochure_IWU.pdf}

\bibitem{Paetz2013}
A.-G. Paetz, T.~Kaschub, P.~Jochem, W.~Fichtner, Load-shifting potentials in households including electric mobility - a comparison of user behaviour with modelling results, in: 2013 10th International Conference on the European Energy Market (EEM), 2013, pp. 1--7.
\newblock \href {https://doi.org/https://doi.org/10.1109/EEM.2013.6607324} {\path{doi:https://doi.org/10.1109/EEM.2013.6607324}}.

\bibitem{Deb2002}
K.~Deb, A.~Pratap, S.~Agarwal, T.~Meyarivan, A fast and elitist multiobjective genetic algorithm: Nsga-ii, IEEE Transactions on Evolutionary Computation 6~(2) (2002) 182--197.
\newblock \href {https://doi.org/https://doi.org/10.1109/4235.996017} {\path{doi:https://doi.org/10.1109/4235.996017}}.

\bibitem{zitzler2001spea2}
E.~Zitzler, M.~Laumanns, L.~Thiele, \href{https://www.research-collection.ethz.ch/handle/20.500.11850/145755}{Spea2: Improving the strength pareto evolutionary algorithm}, TIK report 103 (2001).
\newline\urlprefix\url{https://www.research-collection.ethz.ch/handle/20.500.11850/145755}

\bibitem{Ishibuchi2015}
H.~Ishibuchi, H.~Masuda, Y.~Tanigaki, Y.~Nojima, Modified distance calculation in generational distance and inverted generational distance, in: A.~Gaspar-Cunha, C.~Henggeler~Antunes, C.~C. Coello (Eds.), Evolutionary Multi-Criterion Optimization, Springer International Publishing, Cham, 2015, pp. 110--125.
\newblock \href {https://doi.org/https://doi.org/10.1007/978-3-319-15892-1_8} {\path{doi:https://doi.org/10.1007/978-3-319-15892-1_8}}.

\bibitem{Bader2011}
J.~Bader, E.~Zitzler, {HypE: An Algorithm for Fast Hypervolume-Based Many-Objective Optimization}, Evolutionary Computation 19~(1) (2011) 45--76.
\newblock \href {https://doi.org/https://doi.org/10.1162/EVCO_a_00009} {\path{doi:https://doi.org/10.1162/EVCO_a_00009}}.

\bibitem{gurobi}
{Gurobi Optimization, LLC}, \href{https://www.gurobi.com}{{Gurobi Optimizer Reference Manual}} (2023).
\newline\urlprefix\url{https://www.gurobi.com}

\bibitem{hart2011pyomo}
W.~E. Hart, J.-P. Watson, D.~L. Woodruff, Pyomo: modeling and solving mathematical programs in python, Mathematical Programming Computation 3~(3) (2011) 219--260.
\newblock \href {https://doi.org/https://doi.org/10.1007/s12532-011-0026-8} {\path{doi:https://doi.org/10.1007/s12532-011-0026-8}}.

\bibitem{Pymoo}
J.~Blank, K.~Deb, Pymoo: Multi-objective optimization in python, IEEE Access 8 (2020) 89497--89509.
\newblock \href {https://doi.org/https://doi.org/10.1109/ACCESS.2020.2990567} {\path{doi:https://doi.org/10.1109/ACCESS.2020.2990567}}.

\bibitem{stable-baselines3}
A.~Raffin, A.~Hill, A.~Gleave, A.~Kanervisto, M.~Ernestus, N.~Dormann, \href{http://jmlr.org/papers/v22/20-1364.html}{Stable-baselines3: Reliable reinforcement learning implementations}, Journal of Machine Learning Research 22~(268) (2021) 1--8.
\newline\urlprefix\url{http://jmlr.org/papers/v22/20-1364.html}

\bibitem{towers_gymnasium_2023}
M.~Towers, J.~K. Terry, A.~Kwiatkowski, J.~U. Balis, G.~d. Cola, T.~Deleu, M.~Goulão, A.~Kallinteris, A.~KG, M.~Krimmel, R.~Perez-Vicente, A.~Pierré, S.~Schulhoff, J.~J. Tai, A.~T.~J. Shen, O.~G. Younis, \href{https://zenodo.org/record/8127025}{Gymnasium} (Mar. 2023).
\newblock \href {https://doi.org/10.5281/zenodo.8127026} {\path{doi:10.5281/zenodo.8127026}}.
\newline\urlprefix\url{https://zenodo.org/record/8127025}

\bibitem{NETO20082169}
A.~H. Neto, F.~A.~S. Fiorelli, \href{https://www.sciencedirect.com/science/article/pii/S0378778808001448}{Comparison between detailed model simulation and artificial neural network for forecasting building energy consumption}, Energy and Buildings 40~(12) (2008) 2169--2176.
\newblock \href {https://doi.org/https://doi.org/10.1016/j.enbuild.2008.06.013} {\path{doi:https://doi.org/10.1016/j.enbuild.2008.06.013}}.
\newline\urlprefix\url{https://www.sciencedirect.com/science/article/pii/S0378778808001448}

\bibitem{DENGIZ2021119984}
T.~Dengiz, P.~Jochem, W.~Fichtner, Demand response through decentralized optimization in residential areas with wind and photovoltaics, Energy 223 (2021) 119984.
\newblock \href {https://doi.org/https://doi.org/10.1016/j.energy.2021.119984} {\path{doi:https://doi.org/10.1016/j.energy.2021.119984}}.

\end{thebibliography}
